\documentclass[%
aps,
 twocolumn,prd,
showpacs,preprintnumbers,
 amsmath,amssymb,
 aps,
superscriptaddress]{revtex4-1}

\usepackage{graphicx}
\usepackage{dcolumn}
\usepackage{bm}
\usepackage{hyperref}

\begin{document}

\title{Impact of gravitational radiation higher order modes on single aligned-spin gravitational wave searches for binary black holes.}

\affiliation{Departament de Física \& IAC3, Universitat de les Illes Balears and Institut d'Estudis Espacials de Catalunya,
Cra. Valldemossa km. 7.5, E-07122 Palma de Mallorca, Spain}
\affiliation{ Center for Relativistic Astrophysics and School of Physics
Georgia Institute of Technology, Atlanta, GA 30332}
\affiliation{School of Physics and Astronomy, Cardiff University, Queens Building, CF24 3AA Cardiff, United Kingdom}
\affiliation{Albert-Einstein-Institut, Max-Planck-Institut f\"ur Gravitationsphysik, D-14476 Golm, Germany}
\affiliation{International Centre for Theoretical Sciences, Tata Institute of Fundamental Research, IISc Campus, Bangalore 560012, India}
\author{Juan Calder\'on~Bustillo$^\text{1,2}$}\noaffiliation
\author{Sascha Husa$^\text{1,5}$}\noaffiliation
\author{Alicia M. Sintes Olives $^\text{1}$}\noaffiliation
\author{Michael P\"urrer$^\text{3,4}$}\noaffiliation

\preprint{ LIGO-P1500184}

\begin{abstract}
Current template-based gravitational wave searches for compact binary coalescences (CBC) use waveform models that neglect the higher order modes content of the gravitational radiation emitted, considering only the quadrupolar $(\ell,|m|)=(2,2)$ modes. We study the effect of such a neglection for the case of aligned-spin CBC searches for equal-spin (and non-spinning) binary black holes in the context of two versions of Advanced LIGO: the upcoming 2015 version, known as early Advanced LIGO (eaLIGO) and its Zero-Detuned High Energy Power version, that we will refer to as Advanced LIGO (AdvLIGO). In addition, we study the case of a non-spinning search for initial LIGO (iLIGO). We do this via computing the effectualness of the aligned-spin SEOBNRv1 ROM waveform family, which only considers quadrupolar modes, towards hybrid post-Newtonian/Numerical Relativity waveforms which contain higher order modes. We find that for all LIGO versions, losses of more than $10\%$ of events occur in the case of AdvLIGO for mass ratio $q\geq6$ and total mass $M \geq 100M_\odot$ due to the neglection of higher modes, this region of the parameter space being larger for eaLIGO and iLIGO. Moreover, while the maximum event loss observed over the explored parameter space for AdvLIGO is of $15\%$ of events, for iLIGO and eaLIGO this increases up to $(39,23)\%$. We find that neglection of higher modes leads to observation-averaged systematic parameter biases towards lower spin, total mass and chirp mass. For completeness, we perform a preliminar, non-exhaustive comparison of systematic biases to statistical errors. We find that, for a given SNR, systematic biases dominate over statistical errors at much lower total mass for eaLIGO than for AdvLIGO.
 \end{abstract}

\maketitle


\section{Introduction}

Compact Binary Coalescences (CBC) are the most promising candidates for a first direct detection of gravitational waves (GW). Starting in September 2015, the next generation of GW detectors, Advanced LIGO  \cite{Abbott:2007kv}, Advanced Virgo\cite{advVIRGO} and KAGRA\cite{Somiya:2011np} will come online with and eventually reach sensitivities $\sim 10$ times higher than the previous one, increasing by a factor of $\sim 10^3$ the volume to which they are sensitive. This generates high expectations for imminent first GW detection\cite{Dominik:2014yma}.  
The core of searches for CBC's is the so called matched filter (MF)\cite{MatchedFilter}.The MF technique allows for GW signals to be extracted from background noise provided that a correct model (waveform in our case) of the expected signal is used as a filter of the incoming signal. Otherwise the filter will be suboptimal and the GW signal could be lost or its parameters misidentified. 
Current GW searches for CBC's implement template banks whose waveforms do only contain the quadrupolar $(\ell,|m|)=(2,2)$ modes of the GW emission, known as quadrupolar waveforms. These neglect the higher order mode (HM) content of the incoming signal. This is justified by the fact that, in the non-precessing case, most of the power emitted by the source is carried by these two modes. 

The goal of this paper is to study the consequences of this neglection in current and future GW searches, both in terms of loss of detections and systematic biases caused in the estimation of the parameters (PE) of the source. We will focus on the case of aligned-spin template banks and non-precessing equal aligned-spin BBH within the mass range $50M_\odot<M<220M_\odot$. As target waveforms, we consider equal aligned-spin systems, and as bank waveforms we use the SEOBNRv1 ROM model \cite{Purrer:2014fza}, which describes the quadrupolar modes of equal aligned spins CBC's using a single effective $\chi$ spin parameter \cite{Santamaria:2010yb}\footnote{$\chi=\frac{M_1\chi_1+M_2\chi_2}{M_1+M_2}$, where $\chi_i$ and $M_i$ are respectively the dimensionless spin and mass of the $i^{'}th$ black hole. Note that for the equal spin case one has $\chi=\chi_1=\chi_2$}. In particular, we choose as targets four non-spinning systems with mass ratio $q=\{3,4,6,8\}$ and four spinning cases:  a $q=1$ system with $\chi=\pm 0.2$ and a $q=3$ system with $\chi=\pm0.5$. The latter correspond to those systems available in the public NR SXS catalogue \cite{SXS} having equal spins for which the HM are the strongest and which lie  within the parameter space covered by SEOBNRv1 ROM model. Also, in Appendix 1 we consider the case of a $q=8$ system with unequal spins $(\chi_1,\chi_2)=(0,-0.5)$ and effective spin $\chi=-0.47$. These are summarized in Table. \ref{injections}. We will consider the case of a SEOBNRv1 ROM template bank including a single effective spin parameter $\chi$ for the case of two Advanced LIGO predicted noise curves: the early version (eaLIGO) \cite{Aasi:2013wya} with a lower frequency cutoff $f_0=30$Hz and the design Zero-Detuned-High-Energy-Power version (AdvLIGO) \cite{advLIGOcurves}, with $f_0=10$Hz. Since no detections were made in initial LIGO (iLIGO) data, for which searches have been performed using a non-spinning template bank \cite{Babak:2012zx}, we will also pay attention to the corresponding sensitivity curve \cite{LIGO:2012aa}, for which we will consider non-spinning targets, a non spinning template bank and $f_0=30$Hz. 

The case of non-spinning targets and a non-spinning template bank for the case of AdvLIGO has been widely studied. Pekowsky et. al., \cite{Pekowsky:2012sr} explored the mass range $M>100M_\odot$ and noted that the match between BBH NR waveforms including HM and the corresponding ones including only quadrupolar modes is $<0.97$ for most of the orientations of the binary. They also noticed that however, these orientations coincide with those for which the SNR is the lowest, mitigating the effect of HM when average over orientations is considered.  More recently, Brown et.al.,  \cite{Brown:2012nn} and  Capano et. al., \cite{Capano:2013raa} studied respectively the \textit{fitting factor} (FF) \cite{Apostolatos:1995pj} of a non-spinning quadrupolar template bank towards non-spinning waveforms including HM for the total mass range $m_1,m_2\leq 25 M_{\odot}$ and $m_1,m_2\leq 200M_{\odot}$. The result is that for total masses $M<50_\odot$ and mass ratios $q<4$ one does not expect event losses larger than $10\%$, which is within the commonly accepted limit in GW searches. Furthermore, \cite{Capano:2013raa} also computed the $\chi^2$ \cite{Allen:2004gu} and $\rho_{new}$ \cite{Babak:2012zx} of their target signals towards their non-spinning bank, simuating the effect of HM in a full search neglecting HM and estimated the false alarm rate (FAR) of a search including them. This allowed them to compare the sensitivity of both searches to signals including HM. They concluded that inclusion of HM in current template banks would only be advantageous for certain regions of the parameter space for which the FF of the bank towards their target signals were particularly low. In particular for $q\geq 4$ and $M>100M_\odot$. These event loss results widely agree with those presented by Varma et al., \cite{Varma:2014jxa}, who also studied the systematic parameter bias caused by the neglection of HM and compared it with the statistical uncertainty due to the presence of Gaussian noise in the data stream. They concluded that the former dominate the latter for mass ratio $q\geq4$ and total masses $M>150M_\odot$ for a SNR $\rho \sim 8$.  This study was based on the Fisher information matrix formalism, which allowed them to study a large number of points in the parameter space. In contrast Littenberg et. al., \cite{Littenberg:2012uj} studied the presence of systematic biases in the estimated parameters of the CBC but compared them against the expected statistical errors using Markov-Chain Monte-Carlo (MCMC) techniques.  However, the large computational cost of this study  restricted it to a few points of the parameter space. They obtained that, for binaries such that $1\leq q \leq 6$ and $M < 60M_\odot$ and fixing inclination angle to $\theta = \pi/3$, systematic errors introduced by the neglection of HM are smaller than the expected statistical errors at SNR $\sim 12$. However, for larger masses ($M = 120M_\odot$, $q $= 6), systematic biases will dominate statistical errors at SNR $\sim 12$. Finally, during the preparation of this paper we became aware that Graff et. al., \cite{Graff:2015bba} shown that higher modes are required for parameter estimation and detection of non-spinning high-mass binaries with an SNR $\geq 9$. 

While the above summarized work has considered non-spinning searches and the design Zero-Detuned High Energy Power Advanced LIGO sensitivity curve (AdvLIGO) \cite{advLIGOcurves}, we extend their studies to the case of aligned spin searches \cite{Canton:2014ena} for the early 2015 Advanced LIGO (eaLIGO) sensitivity curve and \cite{Aasi:2013wya}. We also revisit  the case of the initial LIGO (iLIGO) \cite{LIGO:2012aa} sensitivity curve using a non-spinning template bank and  targets. There are various reasons that motivate these choices: the first is that it is expected that aligned-spin searches like \cite{Canton:2014ena} will be implemented in the upcoming Advanced LIGO science runs. The extra degree of freedom that the spin parameter $\chi$ provides could reduce the losses observed for non-spinning targets due to the neglection of HM when non-spinning template banks are considered and of course, we want to test what the effect for spinning systems is. Also, the different sensitivity curves considered and in particular their different frequency cutoff $f_0$, will translate into very different event losses and parameter biases produced. The fact of including an effective spin parameter $\chi$ in our template bank will lead to lower event losses for non-spinning targets than those found in \cite{Canton:2014ena} and \cite{Varma:2014jxa}, we will pay the price of important biases in the estimated spin. This extends the study of Veitch et al.,\cite{Veitch:2015ela} who concluded that the spin of non-spinning BBH (lacking HM) cannot be accurately measured using a single-effective spin parameter template bank. Finally, we will see that the value of the spin has a secondary effect in the impact of HM compared to that of the total mass and mass ratio. 
\begin{figure}[!ht]
\centering
\includegraphics[width=.95\columnwidth]{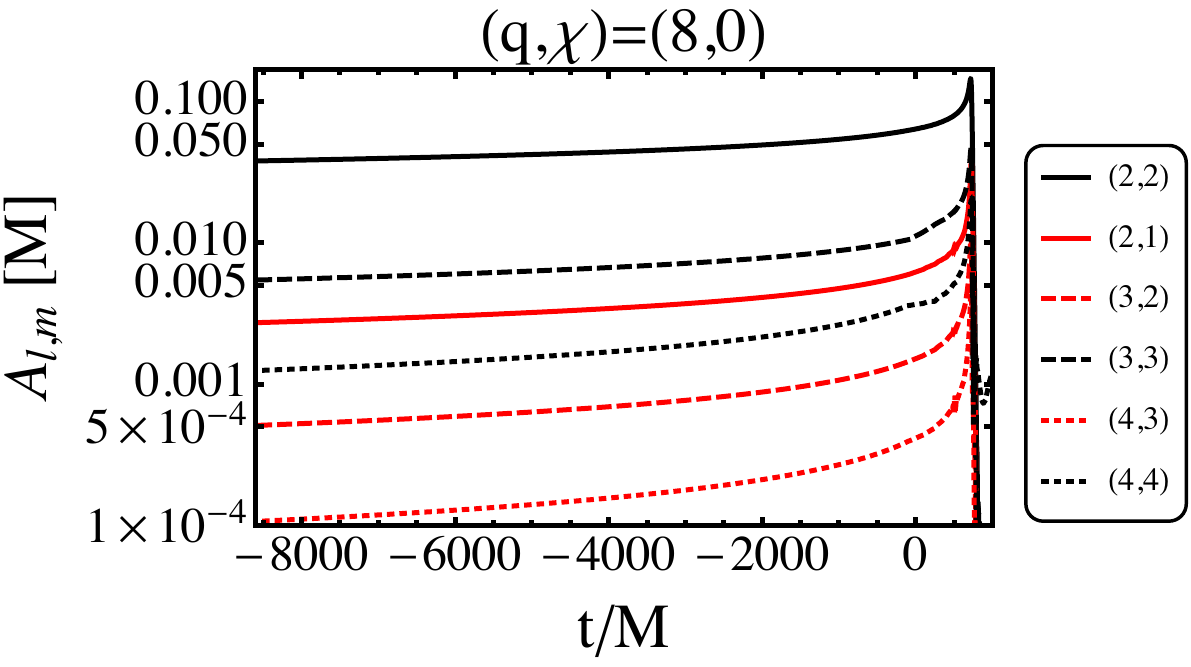}
\caption[ Several modes of a q=8 non-spinning CBC]{Amplitude of the $(\ell,m)$ modes of a $(q,\chi)=(8,0)$ system during the last orbits of the coalescence in logarithmic scale. The modes are the result of hybridizing post-Newtonian Taylor T1 and Numerical Relativity data [see Sec. IV].}
\label{ex:fig:GWhighermodes}
\end{figure} 
\section{Higher Order Modes}
Consider a non-precessing CBC with total mass, mass ratio and effective spin collectively denoted by $\Xi=\{M,q,\chi\}$. Denoting by $d_L$ the luminosity distance between source and detector, consider a frame of reference centered on the source and described by standard spherical coordinates $(d_L,\theta,\varphi)$ such that the $\theta=0$ axis coincides with the total angular momentum of the binary. Then, the strain $h$ produced by an emitted GW with effective polarization $\psi$ \cite{Bustillo:2015ova} at a given point $p$ on its sky can be decomposed as a sum of modes $h_{\ell,m}(\Xi;t)$ weighted by spin -2 weighted spherical harmonics\cite{Sharmonics} $Y^{-2}_{\ell,m}(\theta,\varphi)$ as:
\begin{equation}
\begin{aligned}
&h(\Xi;d_L,\theta,\varphi,\psi;t)\\
&=\frac{F}{d_L}({\cal R} \cos\psi + {\cal I} \sin{\psi})
\sum_{\ell\geq 2}\sum_{m=-\ell}^{m=\ell}Y^{-2}_{\ell,m}(\theta,\varphi)h_{\ell,m}(\Xi;t),
\end{aligned}
\end{equation}\\
where ${\cal R}$ and ${\cal I}$ denote the real and imaginary part operators, $h_{\ell,m}(\Xi;t)=A_{\ell,m}(\Xi;t)e^{-i\phi_{\ell,m}(\Xi;t)}$, $A_{\ell,m}$ being real, and the factor $F$ encodes the amplitude of the antenna pattern of the detector \cite{Jaranowski:1998qm,Varma:2014jxa}.
 Fig.\ref{ex:fig:GWhighermodes} shows the amplitude of the most dominant modes for a non-spinning $q=8$ binary.\\
\begin{figure}[!ht]
\centering
\includegraphics[width=.45\columnwidth]{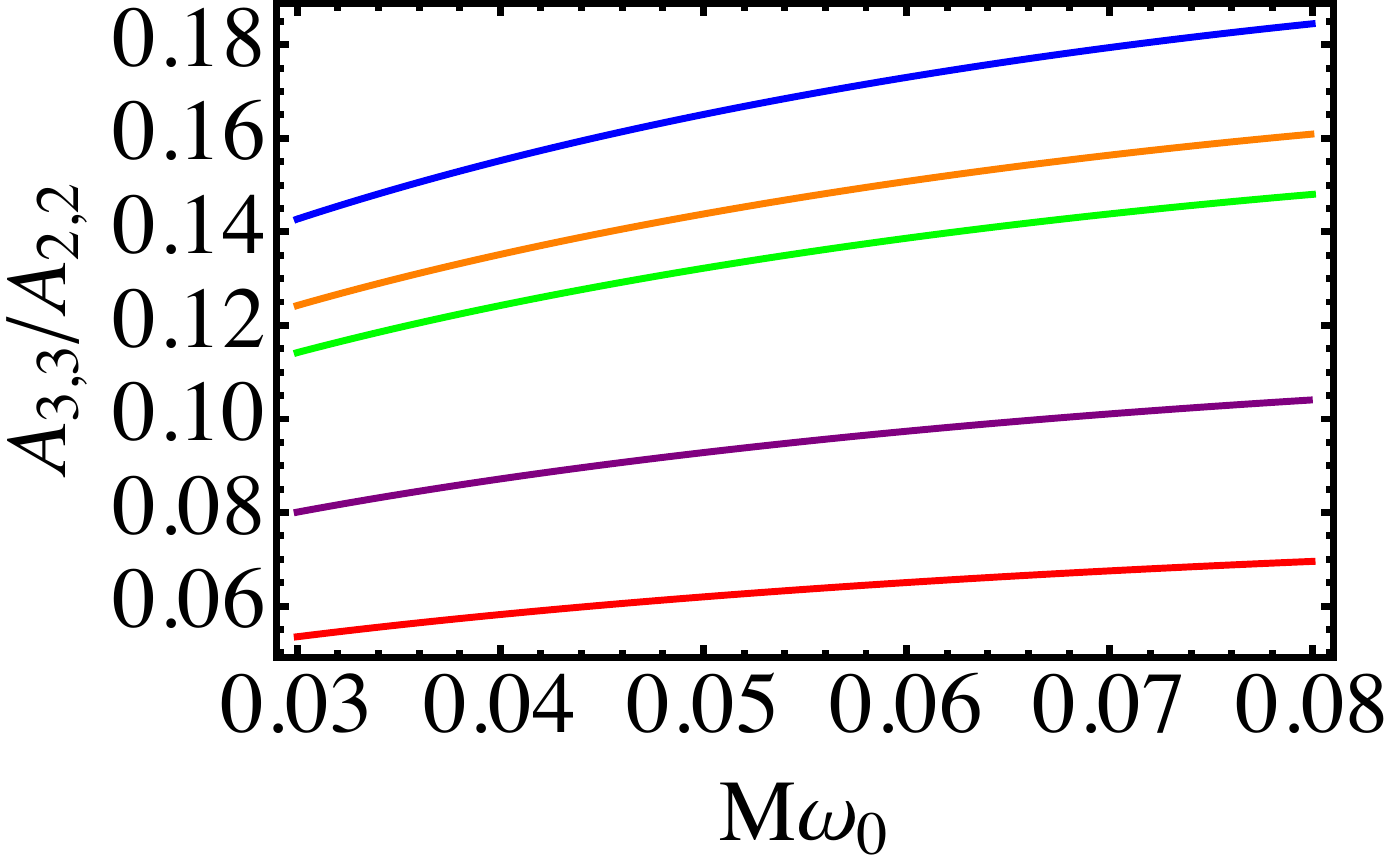}
\includegraphics[width=.53\columnwidth]{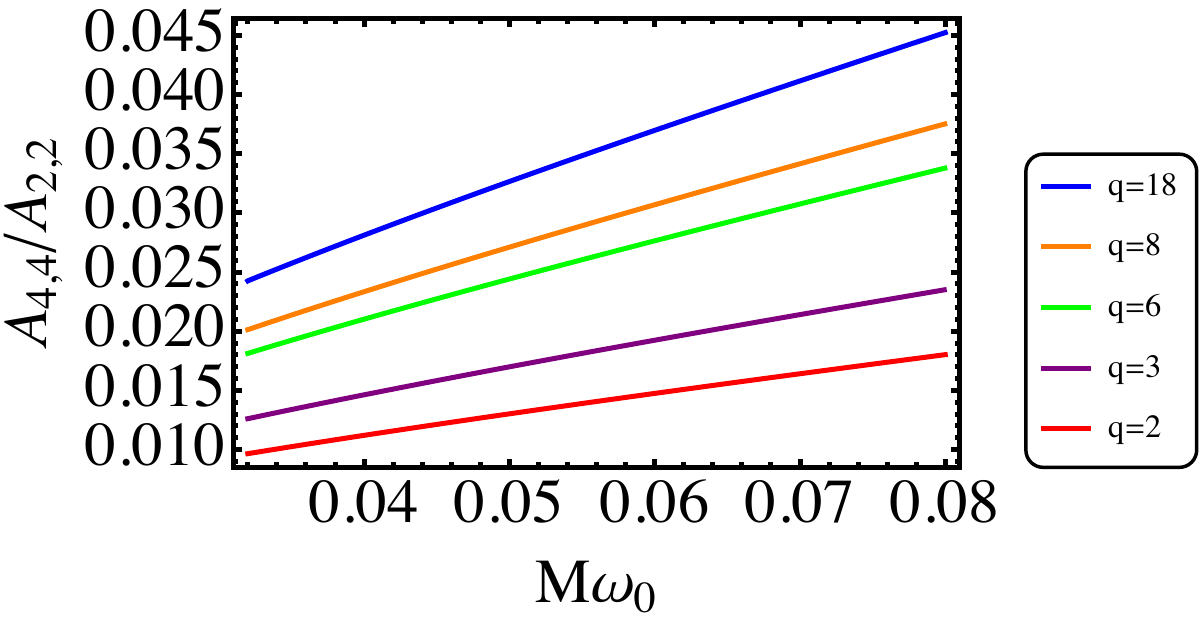}
\includegraphics[width=.45\columnwidth]{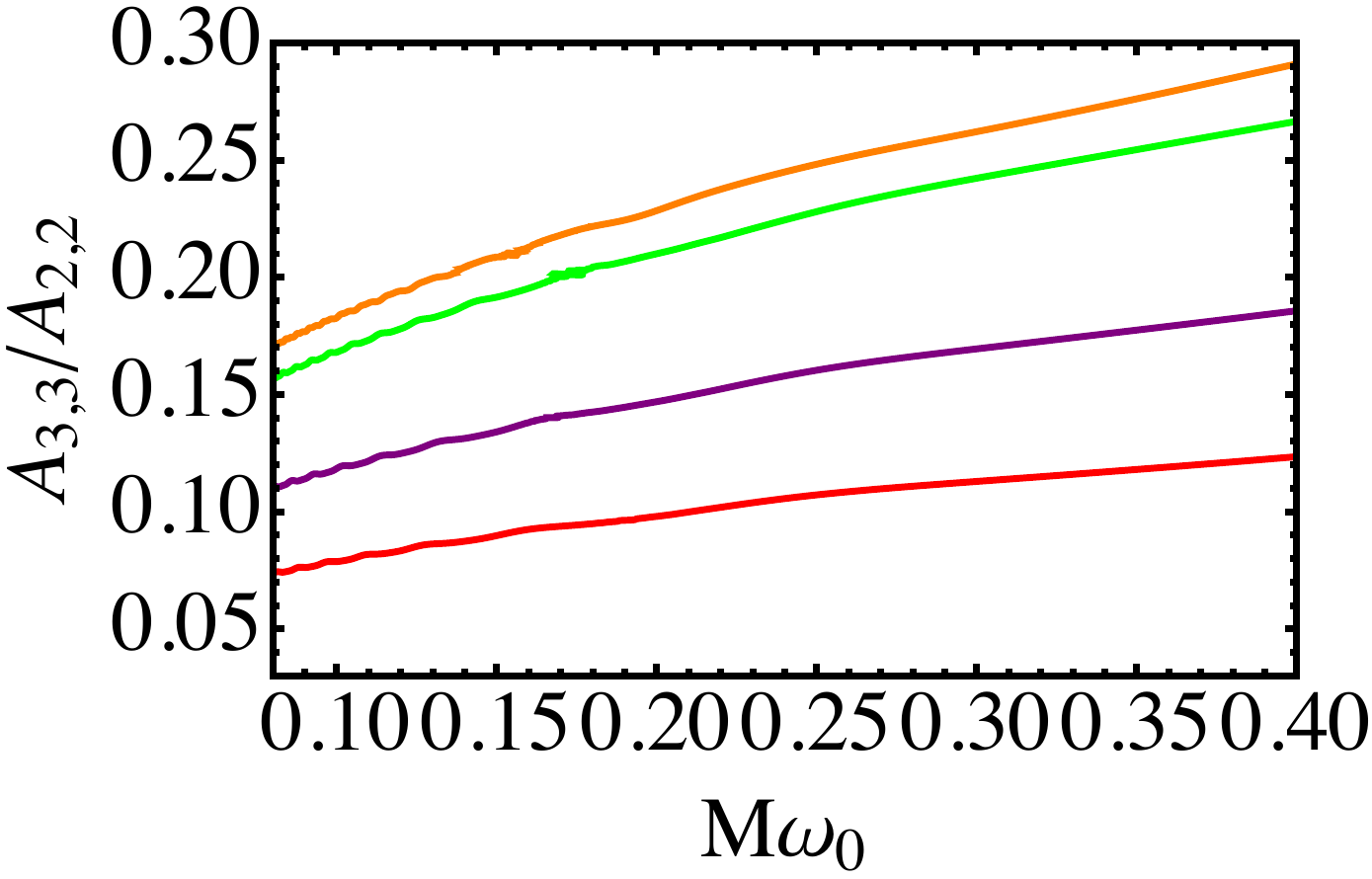}
\includegraphics[width=.53\columnwidth]{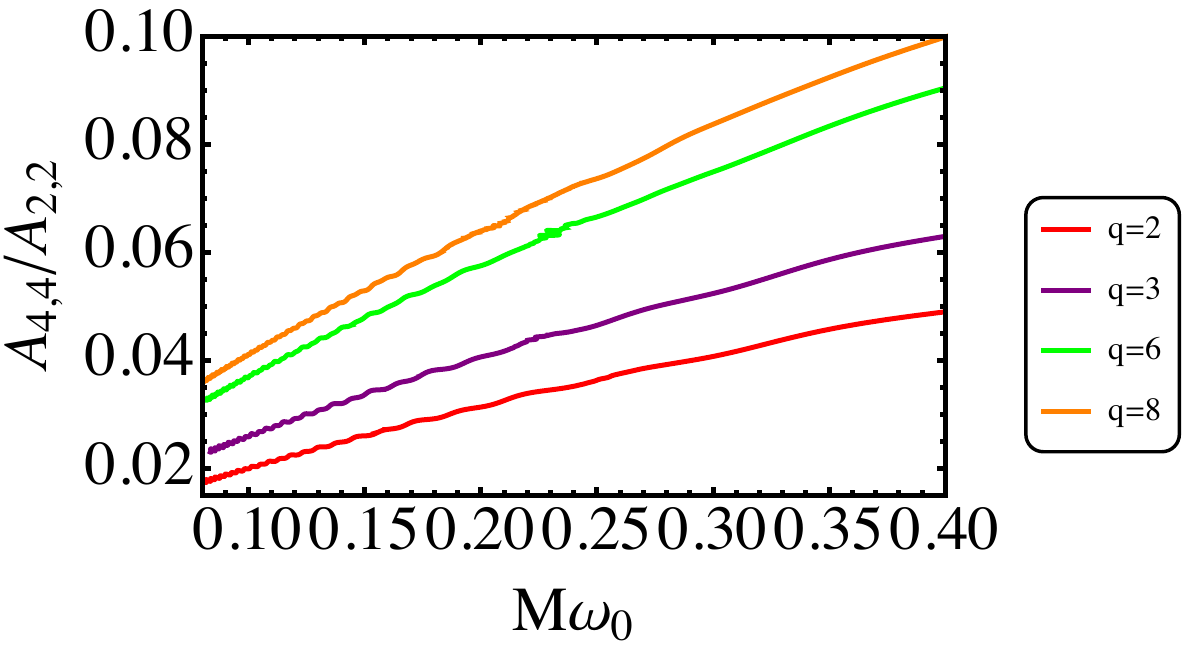}
\caption[Effect of $q$ on higher modes]{Relative T1 (Top) and NR (Bottom) amplitude of the higher modes $(3,3)$ and $(4,4)$ relative to the dominant $(2,2)$ mode as a function of the frequency for several non-spinning systems.  Note how the larger the mass ratio $q$, which increases from bottom to top in the plots, the larger the contribution from higher order modes. As in \cite{Bustillo:2015ova}, we notice the differences between the predicted amplitudes by both PN and NR mainly due to PN truncation error. Note that the merger, taken as coincident with the amplitude maximum of the $(2,2)$ mode, occur between $M\omega_0\simeq 0.30$ for $q=8$ and $M\omega_0\simeq 0.36$ for $q=2$ in the bottom plots.}
\label{ex:fig:hmq}
\end{figure}
The effect that HM have in the observed signal depends on three main factors. First, regarding the source, post-Newtonian results yield that the larger the mass ratio, the larger the ratio $A_{\ell,m} /A_{2,2}$ is \cite{Blanchet2014}, as can be noticed in the top row of Fig.\ref{ex:fig:hmq}, where $\omega_0$ represents the frequency of the $(2,2)$ mode, i.e., $M\omega_0=M d\phi_{2,2}/dt$. Note how in the frequency range shown in these plots, the post-Newtonian amplitude of the $(2,2)$ mode is about 1 order of magnitude larger than that of the next most dominant mode (typically the $(3,3)$, when present) for all the sources shown. However, although this behaviour is qualitatively kept through the late inspiral and merger (at $M\omega_0\simeq 0.33$ in the bottom plots.), the NR amplitudes shown in the bottom row of  Fig.\ref{ex:fig:hmq}, show that this ratio can get up to $\sim 0.3$ for the case of a $q=8$ non-spinning system. As a general trend, the larger $q$ is, the larger the contribution from HM will be. This will translate into larger event losses due to neglection of HM for larger $q$.\\

\begin{figure}[!hb]
\centering
\includegraphics[width=.45\columnwidth]{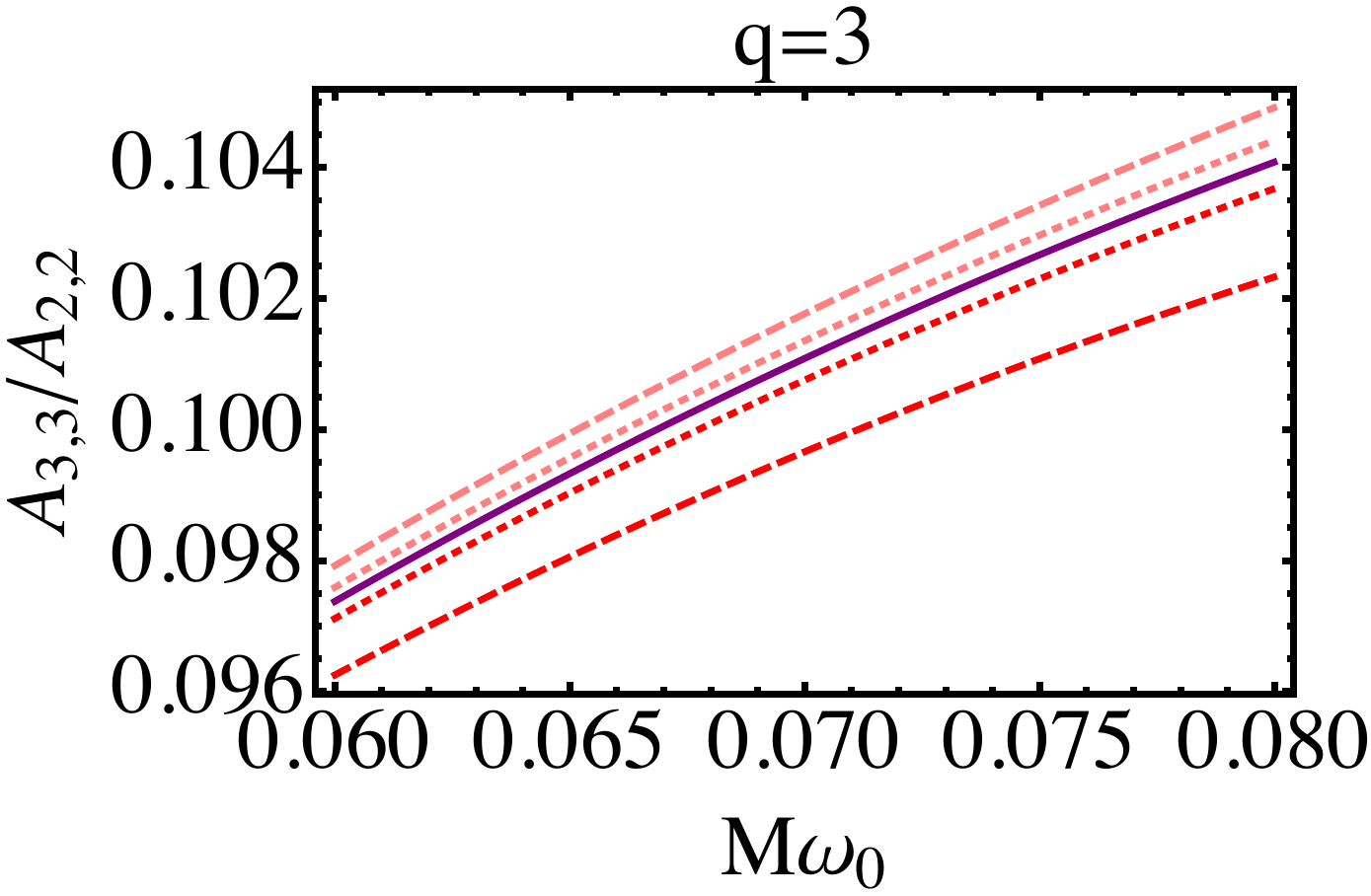}
\includegraphics[width=.53\columnwidth]{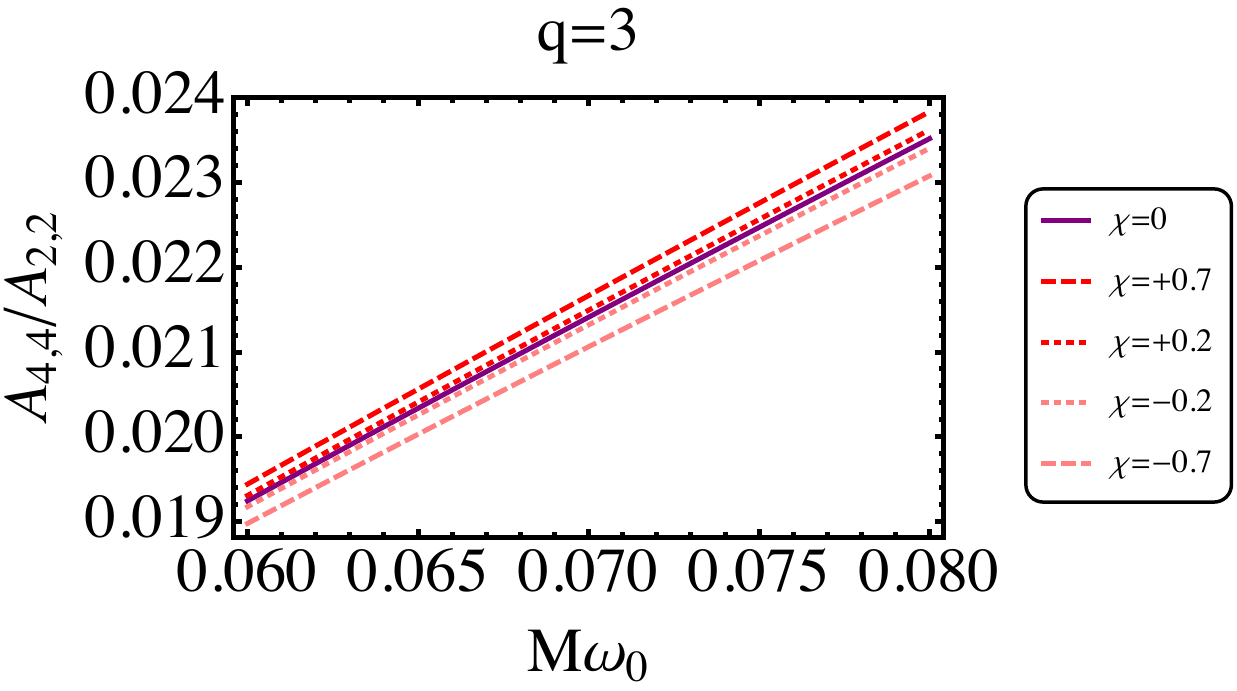}
\includegraphics[width=.45\columnwidth]{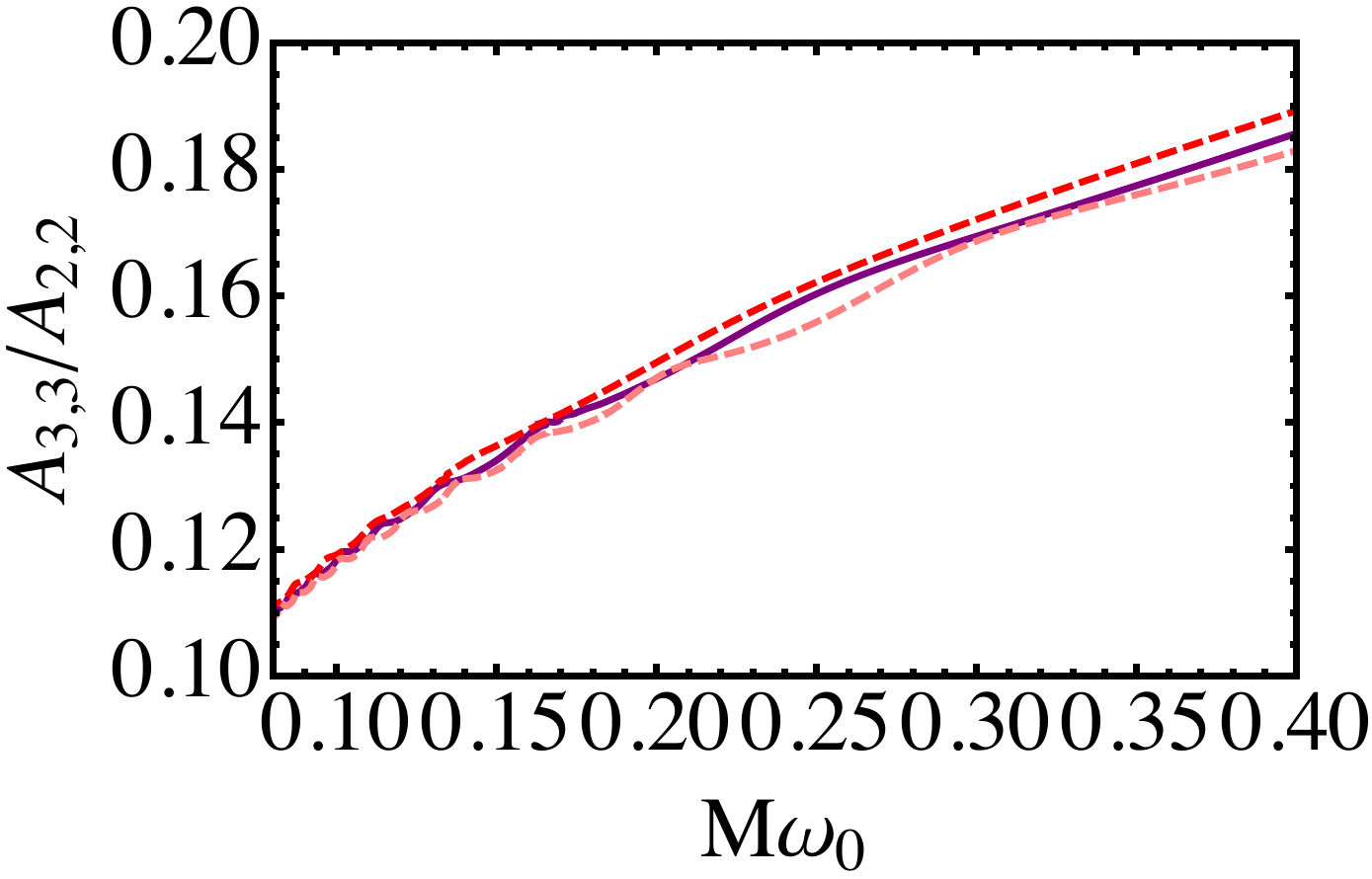}
\includegraphics[width=.53\columnwidth]{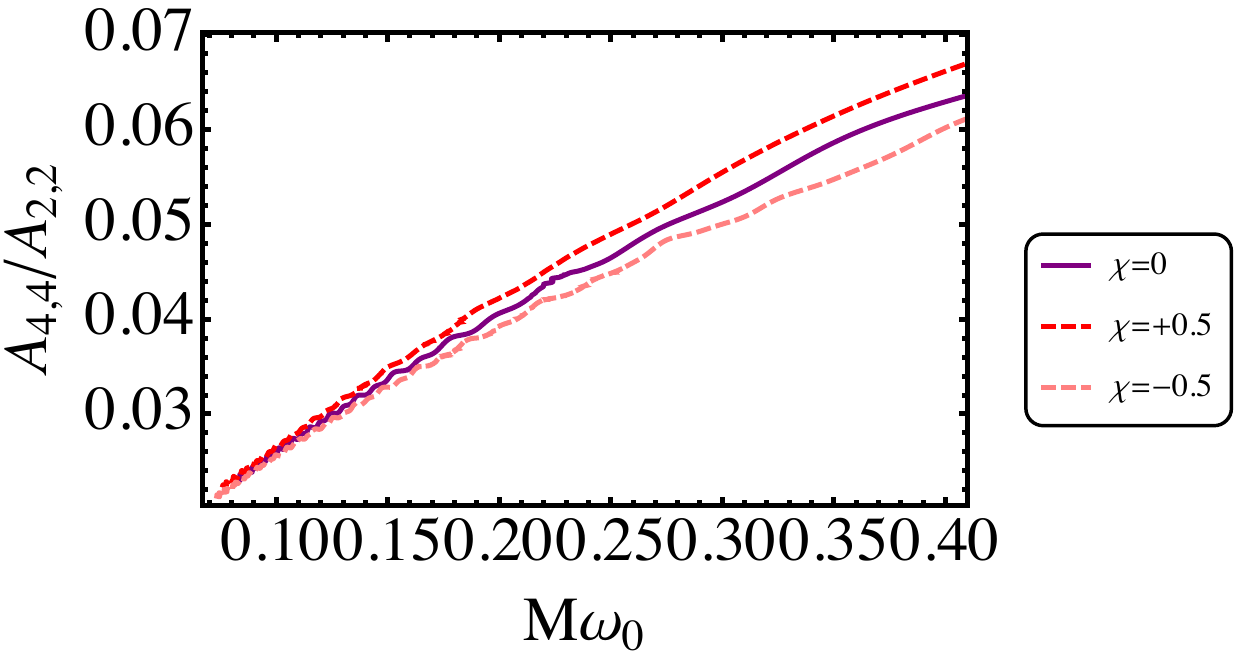}
\caption[Effect of $\chi$ on higher modes]{Relative T1 (Top) and NR (Bottom) amplitude of the $(3,3)$ and $(4,4)$ modes relative to the dominant $(2,2)$ mode as a function of the frequency for several $q=3$ spinning systems.}
\label{ex:fig:hmspin}
\end{figure}

\begin{figure}[!ht]
\centering
\includegraphics[width=\columnwidth]{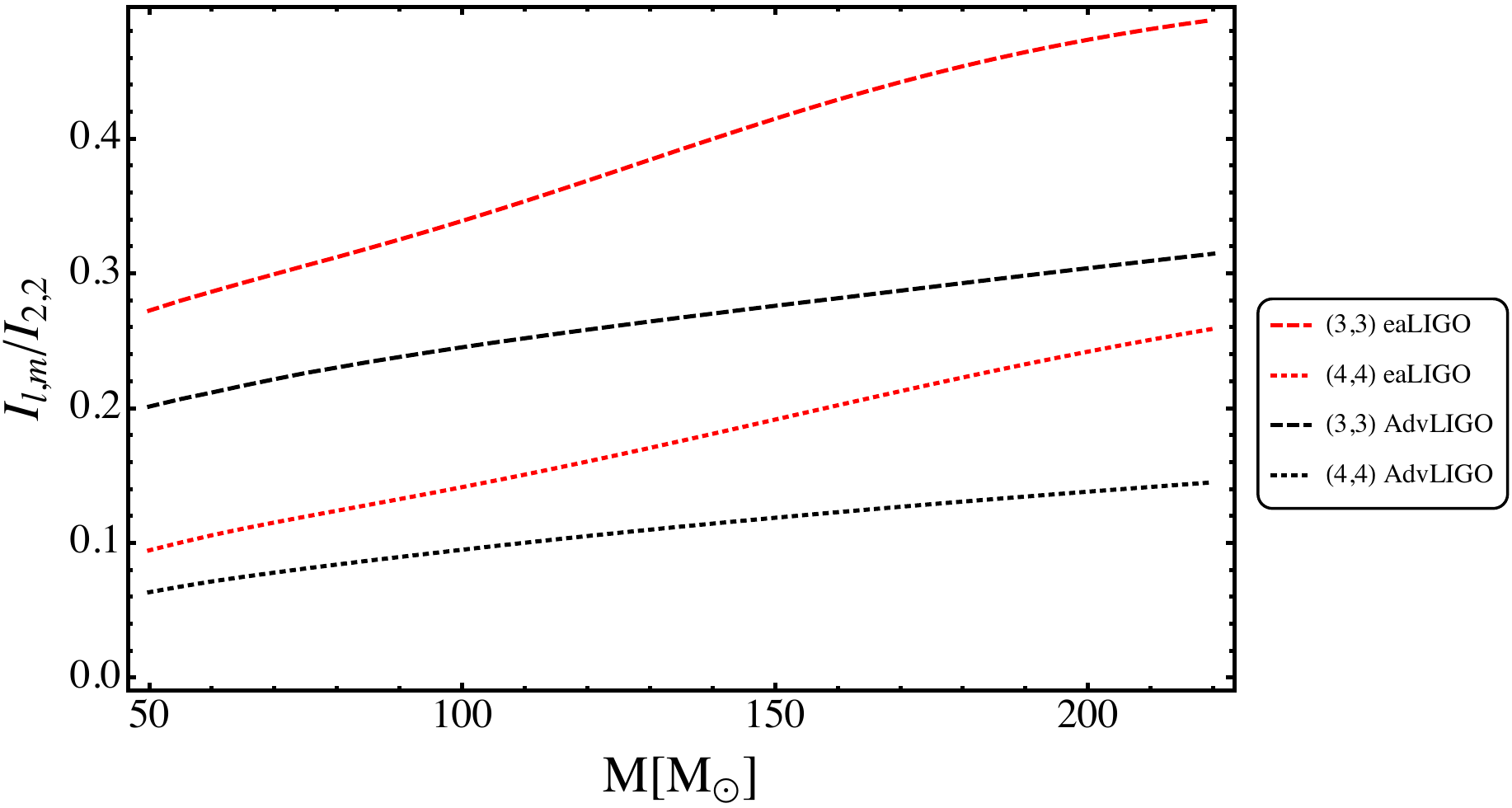}
\caption[Effect of $\chi$ on higher modes]{Value of the ratio $I_{3,3}/I_{2,2}$ and $I_{4,4}/I_{2,2}$ as a function of the total mass of a non-spinning $q=8$ binary for the cases of eaLIGO and AdvLIGO. Note how the $(2,2)$ mode is more dominant for AdvLIGO while the contribution of HM is larger for the case of eaLIGO for all the mass range.}
\label{fig:Ilm}
\end{figure}

The effect of the spin is a bit more intricate since the contribution of the different modes as a function of the spin is mode-dependent. As an example, top and bottom rows of Fig. \ref{ex:fig:hmspin} show respectively the PN and NR ratio between the amplitudes of the $(3,3)$ and $(4,4)$ modes wrt., that of the $(2,2)$: while the relative amplitude of the $(3,3)$ mode grows as the spin gets more positive, the behavior of the $(4,4)$ is the opposite in the PN case. Note however that the range of variation of the ratios shown in these plots is much tinier than that in Fig.\ref{ex:fig:hmq}, which suggests that spin should have a sub-dominant effect compared to that of the mass ratio.
The location of the detector on the sky of the source adds a second factor: the $Y^{-2}_{2,2}$ spherical harmonic is weaker at close to edge-on orientations $(\theta=\pi/2)$, where some higher ones have their maximums. This implies that signals from edge-on systems will have a larger HM content.\\
\begin{center}
\begin{figure*}[!ht]
\includegraphics[width=1.1\columnwidth]{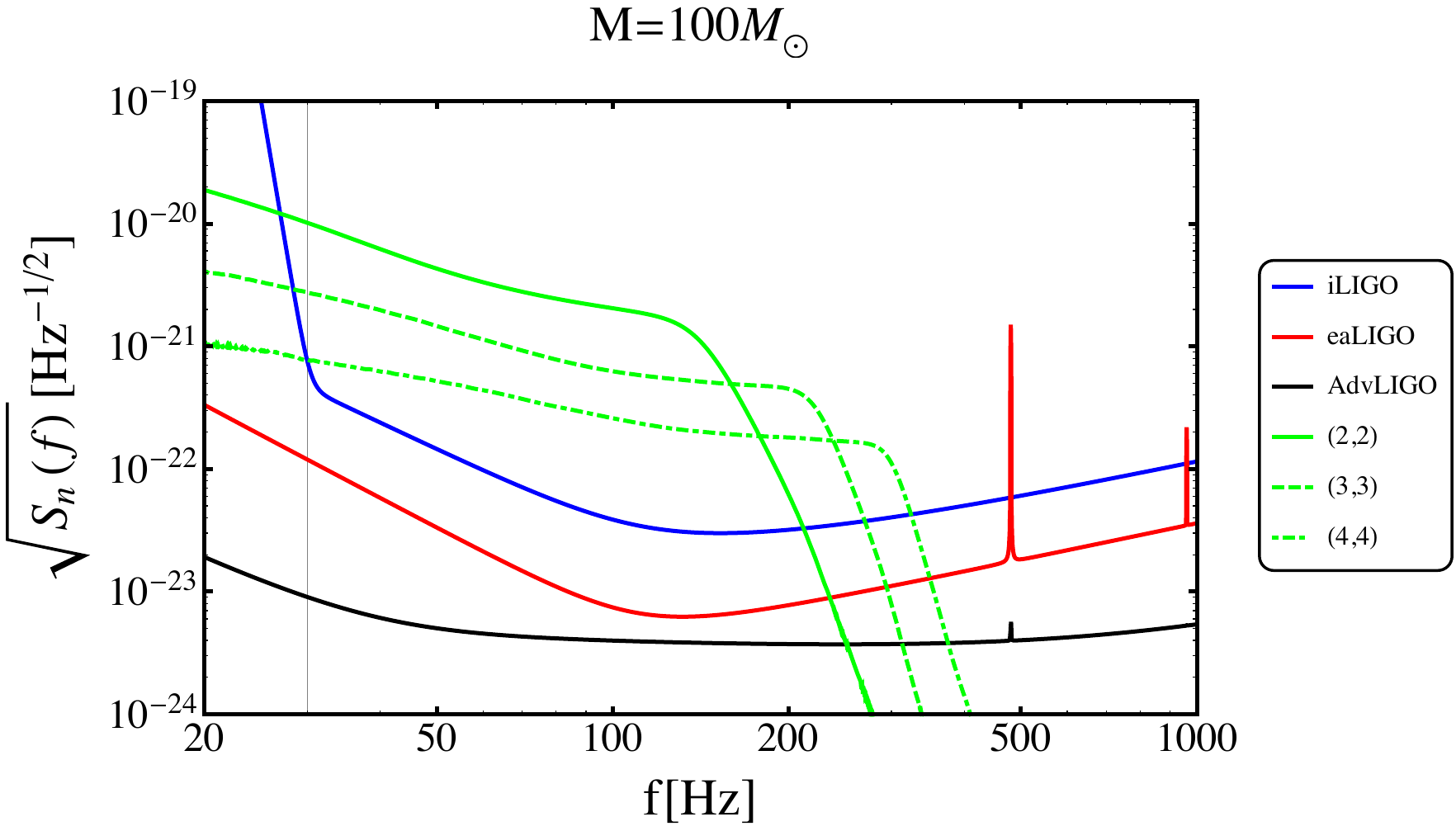}
\includegraphics[width=0.9\columnwidth]{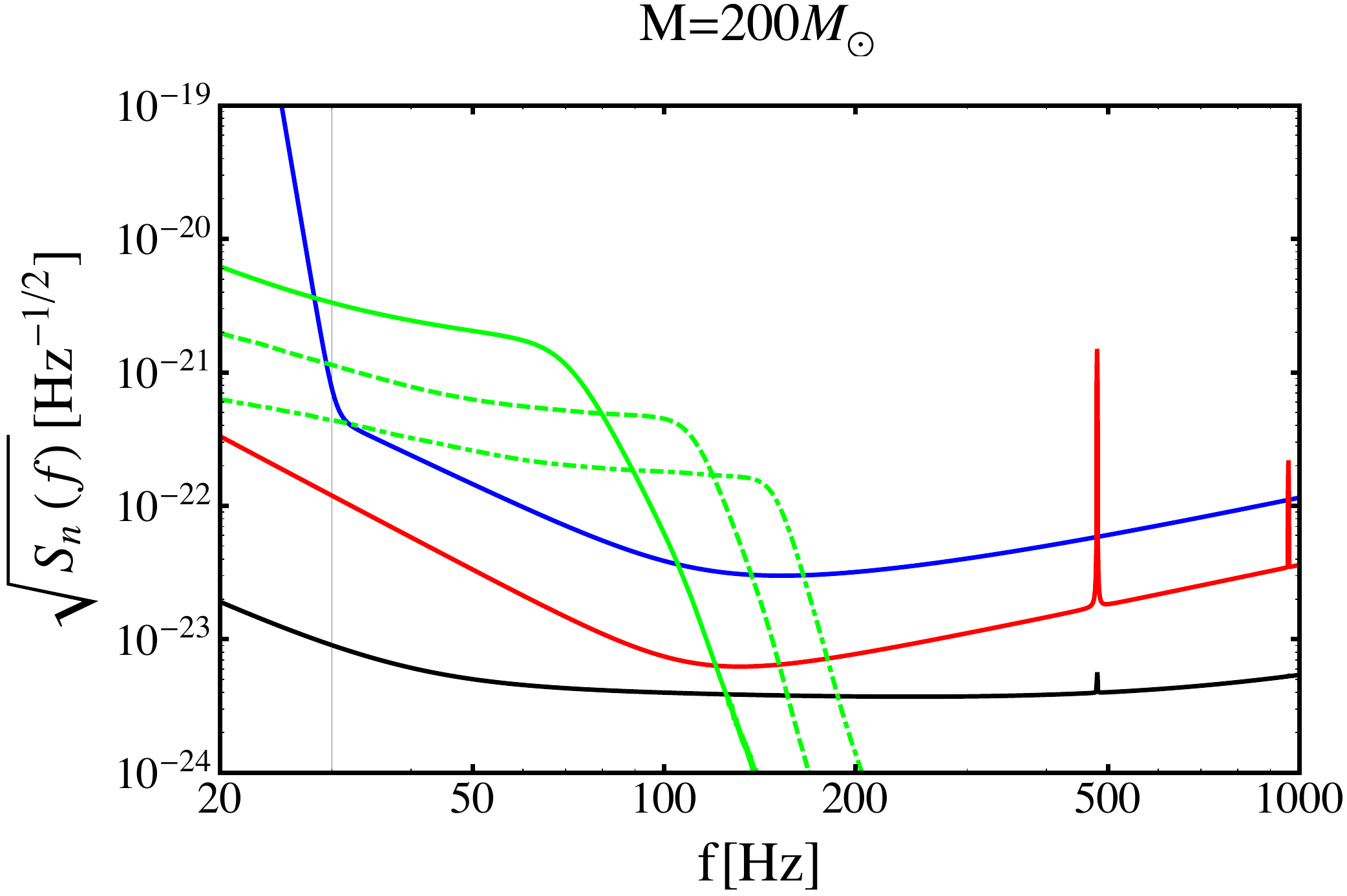}
\includegraphics[width=1.05\columnwidth]{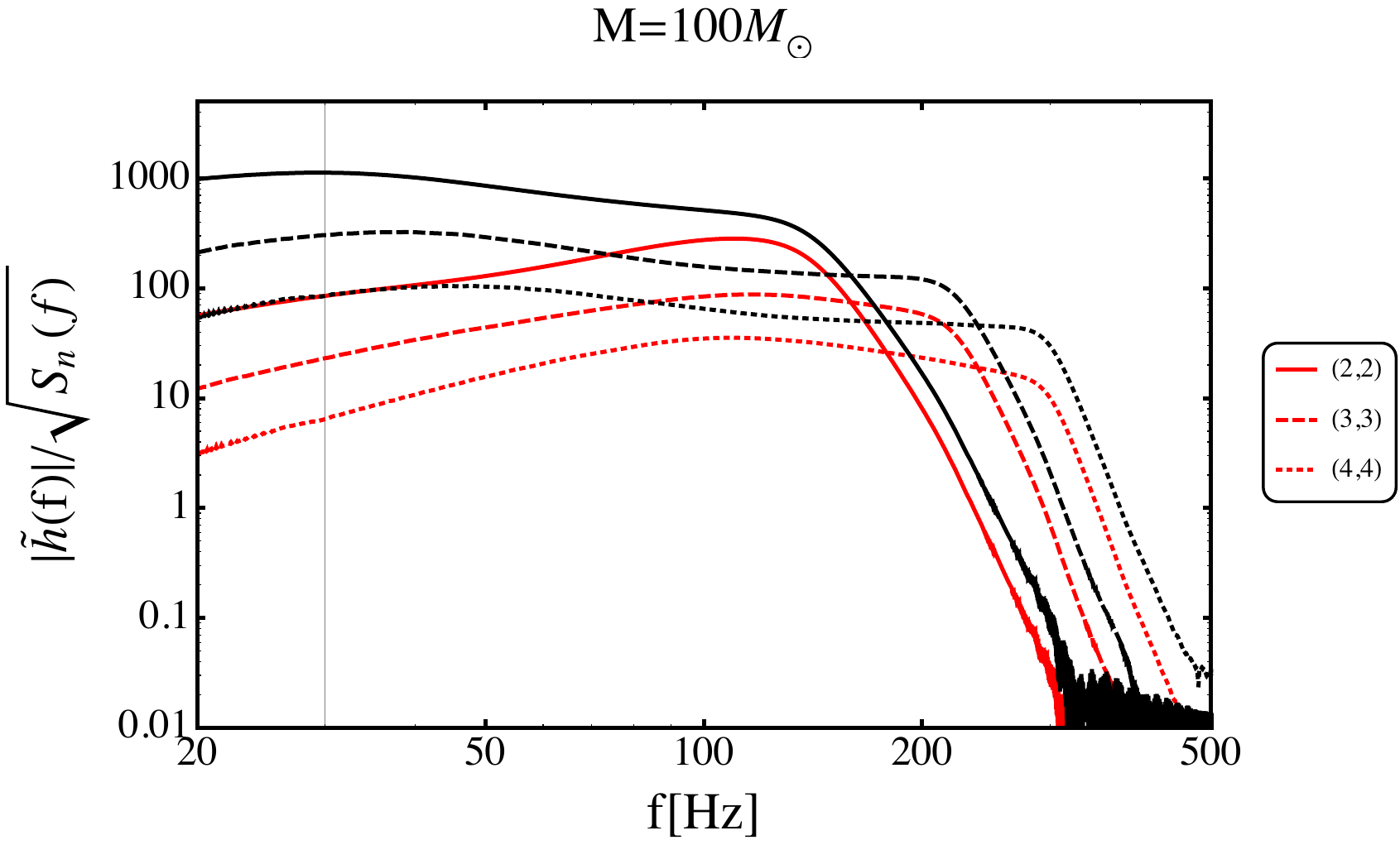}
\includegraphics[width=.95\columnwidth]{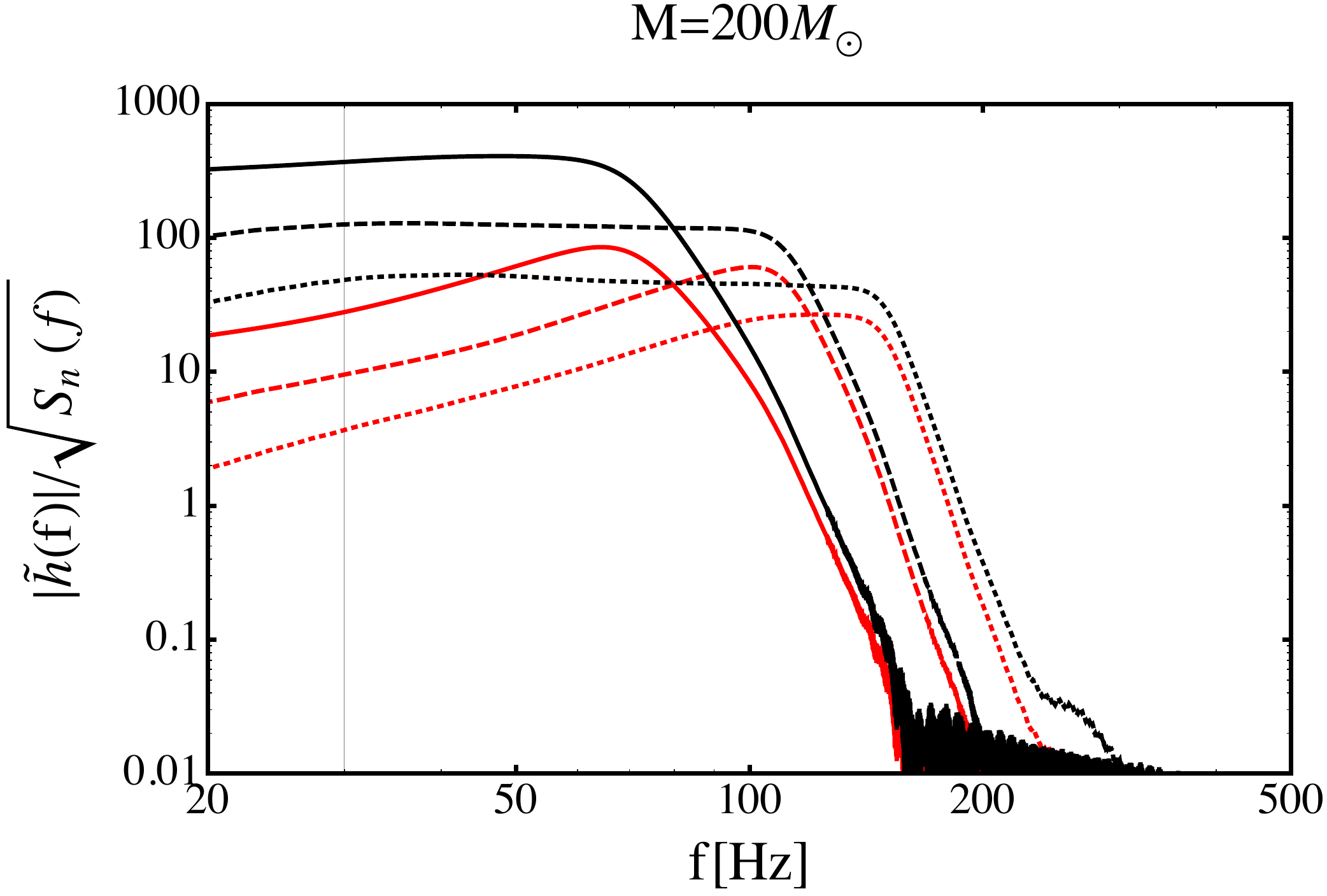}
\caption[Influence of the higher modes in the detector band]{The upper panels show the absolute value of the Fourier transform $\tilde{h}_{\ell,m}(f) \times \Delta f^{1/2}$ of the $(2,2)$, $(3,3)$ and $(4,4)$ modes of a non-spinning $q=8$ binary and the three noise curves considered in this paper. The modes have been re-scaled by an arbitrary factor $(\Delta f^{1/2}/d_L)$ to clearly stand out from the noise curves, since we are only interested in their relative values. The vertical line marks the $30$Hz cutoff of eaLIGO and iLIGO. Note how for the case of $M=100M_\odot$, the $(2,2)$ mode clearly dominates at the sweet-spot of the different noise curves, while this is not the case when $M=200M_\odot$, resulting in a higher contribution of HM in the latter case. The bottom panels show the corresponding whitened templates for eaLIGO(red) and AdvLIGO(black). The lower frequency cutoff of AdvLIGO, together with its flatter sensitivity curve, makes the detector sensitive to a much longer inspiral, clearly dominated by the $(2,2)$ mode, and whose amplitude (unlike for eaLIGO) dominates that of the higher modes peaks corresponding to the merger stage. This makes contribution from HM to be weaker for AdvLIGO. Also, it can be noticed how for the $200M_\odot$ eaLIGO case, the peaks of all whitened templates have similar amplitudes due to the $(2,2)$ mode being clearly out of the sweet-spot while the HM are in, which does not happen for AdvLIGO.}
\label{fig:modes}
\end{figure*}
\end{center}
Finally, as noted in \cite{VanDenBroeck:2006qu}, there is a combined effect of the detector sensitivity curve and the total mass $M$ of the CBC: the frequency of each mode roughly scales with the orbital frequency as $\omega_{\ell,m}(t)=\frac{d\phi_{\ell,m}}{dt}\simeq m\times \omega_{orb}(t)$ and as the total mass $M$ increases, $\omega_{orb}(t)$ falls off as $1/M$. When the total mass $M$ of the source is such that the frequency of the $(2,2)$ mode is below the detector lower frequency cutoff ($f_0$), larger $m$ modes will dominate the incoming signal in band. This will make the observed signal be very different from a quadrupolar waveform. In particular, the lower the seismic wall (the lower the frequency cutoff), the longer PN inspiral (strongly dominated by the $(2,2)$ mode) the detector will be sensitive to. For this reason we decided to study both the cases of AdvLIGO with a $f_0=10$ Hz frequency cutoff and  eaLIGO and iLIGO with $f_0=30$ Hz. As we will see, the different $f_0$ generates notably different event losses. This effect is visualized in Fig. \ref{fig:modes}, where the upper panels show the absolute value of the Fourier transform of the three most dominant modes of a $q=8$ non-spinning binary for the cases of $M=100M_\odot$ and $M=200M_\odot$, and the bottom ones the corresponding whitened versions, $|\tilde{h}_{\ell,m}(f)|/\sqrt{S_n(f)}$, for both eaLIGO and AdvLIGO. Notice here how the larger flatness and lower frequency cutoff of AdvLIGO makes the $(2,2)$ mode clearly dominate in all the plots shown (particularly at the sweet-spot \footnote{Frequencies for which the noise curve reaches its minimum value.} of the noise curve), while for the case of eaLIGO contributions from HM get comparable to that of the $(2,2)$ for high-mass cases.
In order to estimate how important the contribution of HM will be as a function of the total mass and the detector curve, Fig. \ref{fig:Ilm} shows the value of the ratio $I_{\ell,m}/I_{2,2}$, where $I_{\ell,m}=\sqrt{\langle h_{\ell,m}|h_{\ell,m} \rangle}$, as a function of the total mass of the binary. Note how this ratio grows for the case of eaLIGO.

\section{Data Analysis}
Given two real waveforms, $h$ and $g$, and the one sided power spectral density curve $S_n(f)$ of a detector, the inner product $\langle h|g \rangle$ can be expressed as
\begin{eqnarray}
\langle h|g \rangle= 4\Re \int_{f_0}^{\infty}\frac{\tilde{h}(f)\tilde{g}^*(f)}{S_n(f)} df,
\end{eqnarray}
$f_0$ being the lower frequency cutoff of the considered noise curve\footnote{As previously mentioned we consider $f_0=10$Hz for AdvLIGO and $f_0=30$Hz for eaLIGO and iLIGO.}.
The overlap of $h$ and $g$ is then defined as 
\begin{eqnarray}
{\cal O}=\frac{\langle h|g\rangle}{\sqrt{\langle h|h\rangle \langle g|g\rangle}}.
\end{eqnarray}
The signal-to-noise-ratio (SNR) of a signal $s$ when filtered with a template $h$ is then given by
\begin{eqnarray}
\rho=\frac{\langle s| h\rangle}{\sqrt{\langle h|h\rangle}}.
\end{eqnarray}
An output signal $s$ is in general a combination of a GW signal $g$ and background noise $n$. If one assumes the background noise to be Gaussian and with zero-mean, as we will do in this paper, the SNR is directly related to the probability that $h$ is buried in $s$ and to the distance at which it can be detected.
Let us denote $(\theta,\varphi,\psi)\equiv \Lambda$. We then define the match ${\cal M}_h g$ as the overlap ${\cal O}(h|g)$ maximized over relative time-shifts and the extrinsic parameters of $g$, $\Lambda_g$. The fitting factor (or effectualness) $\cal F$ of a bank ${\cal B}$ containing waveforms $h^{\cal B}_i$ with intrinsic parameters $\Xi_i^{\cal B}$ to a waveform $h$ is then defined as \cite{Apostolatos:1995pj}
\begin{eqnarray}
{\cal F}_{\cal B} h=\max_{i}{\cal M}_h h^{\cal B}_i(\Xi_i^{\cal B})
\end{eqnarray}
and represents the fraction of SNR that the bank $\cal B$ can recover from the waveform $h$ at the cost, in general, of a bias $\delta \Xi$ in the estimation of the intrinsic parameters $\Xi$ of $h$. This is, if $h^{\cal B}$ is the waveform of $\cal B$ which has the best overlap with $h$, then in general, $\Xi^{\cal B}=\Xi+\delta \Xi$. 
\section{Analysis set up}
We use as target signals hybrid PN/NR waveforms containing HM as built in \cite{Bustillo:2015ova}. The early inspiral part of the hybrids is built post-Newtonian data computed via the TaylorT1 approximant including 3.5 PN non-spinning \cite{Blanchet2014} and spin-orbit \cite{Bohe2013} and 2PN spin-spin \cite{Mikoczi2005} phase corrections. We include 3PN non-spinning amplitude corrections for the HM \cite{Blanchet2008} and 3.5PN for the 22 mode \cite{Faye2012}. Spin corrections to the amplitudes are used up to 2PN \cite{Buonanno2013}. The late inspiral and merger are described by NR waveforms extrapolated to null infinity to polynomial order $N=2$. The latter have been obtained from the publicly available SXS catalogue \cite{SXS,Szilagyi:2009qz,Scheel:2008rj,Boyle:2009vi}. The $(2,2)$ mode of all target waveforms starts at 10Hz for $M=45M_\odot$. The cases $q\neq 1$ included the $\{2\pm1,2\pm2,3\pm2,3\pm3,4\pm3,4\pm4\}$ modes while $q=1$ cases included the $\{2\pm2,3\pm2 ,4\pm 4\}$ modes.\\
For each hybrid waveform $h$ in Table \ref{injections} we construct all the signals $h_{i,j}(\Xi_i,\Lambda_j)$ for all the values of $M$ and $\Lambda$ in Table.~\ref{grid}
\begin{table}
    \begin{tabular}{|l|l|l|l|l|}
  \hline
    SIM ID & q & $\chi$ & PN & $M\omega_{hyb}$ \\ \hline
   
    SXS:BBH:0168  & 3 & 0                 & T1 & 0.043           \\ \hline
    SXS:BBH:0167   & 4 & 0                & T1  & 0.045               \\ \hline
    SXS:BBH:0166   & 6 & 0                & T1  & 0.045                \\ \hline
    SXS:BBH:0063   & 8 & 0                & T1  & 0.043                \\ \hline
    SXS:BBH:0150  & 1 & +0.2           & T1  & 0.035              \\ \hline
    SXS:BBH:0149  & 1 & -0.2            & T1 & 0.043              \\ \hline
    SXS:BBH:0046   & 3 & +0.5          & T1  & 0.038                \\ \hline
    SXS:BBH:0047   & 3 & -0.5           & T1  & 0.043               \\ \hline
    SXS:BBH:0064   & 8 & -0.47           & only NR  & 0.042               \\ \hline
    \end{tabular}
    \caption{Summary of hybrid waveforms used as target waveforms. If one expresses $h_{2,2}=A_{2,2}e^{i\phi_{2,2}}$, $A_{2,2}$ being real, $M\omega_{hyb}=M d\phi_{2,2}/dt$ indicates the hybridization frequency of the $(2,2)$ mode.}
    \label{injections}
\end{table}
\begin{table}[h]
\begin{tabular}{|l|c|c|c|c|}
\hline
\textbf{Magnitude} & $M$         & \textbf{$cos{\theta}$} & \textbf{$\varphi$} & \textbf{$\psi$} \\ \hline
\textbf{Range}     & [50,218]$M_{\odot}$ & $[0,1]$                & $[0,2\pi)$      & $[0,\pi)$         \\ \hline
\textbf{Step}      & 12$M_{\odot}$       & 0.05                   & $\pi/20$        & $\pi/6$           \\ \hline
\end{tabular}

       \caption{Grid in Mass and angles $\Lambda$ used for our studies.}
       \label{grid}
\end{table}
The described grid suffices for describing all the possible $(\theta,\varphi,\psi)$ since in the non-precessing case it holds 
\begin{eqnarray}
\begin{aligned}
h(\pi-\theta,\varphi,\psi)&=h(\theta,\varphi,\pi-\psi) \\
h(\theta,\varphi,\pi+\psi)&=-h(\theta,\varphi,\psi).
\end{aligned}
\end{eqnarray}

\begin{figure*}[!ht]
\centering
\includegraphics[width=1\columnwidth]{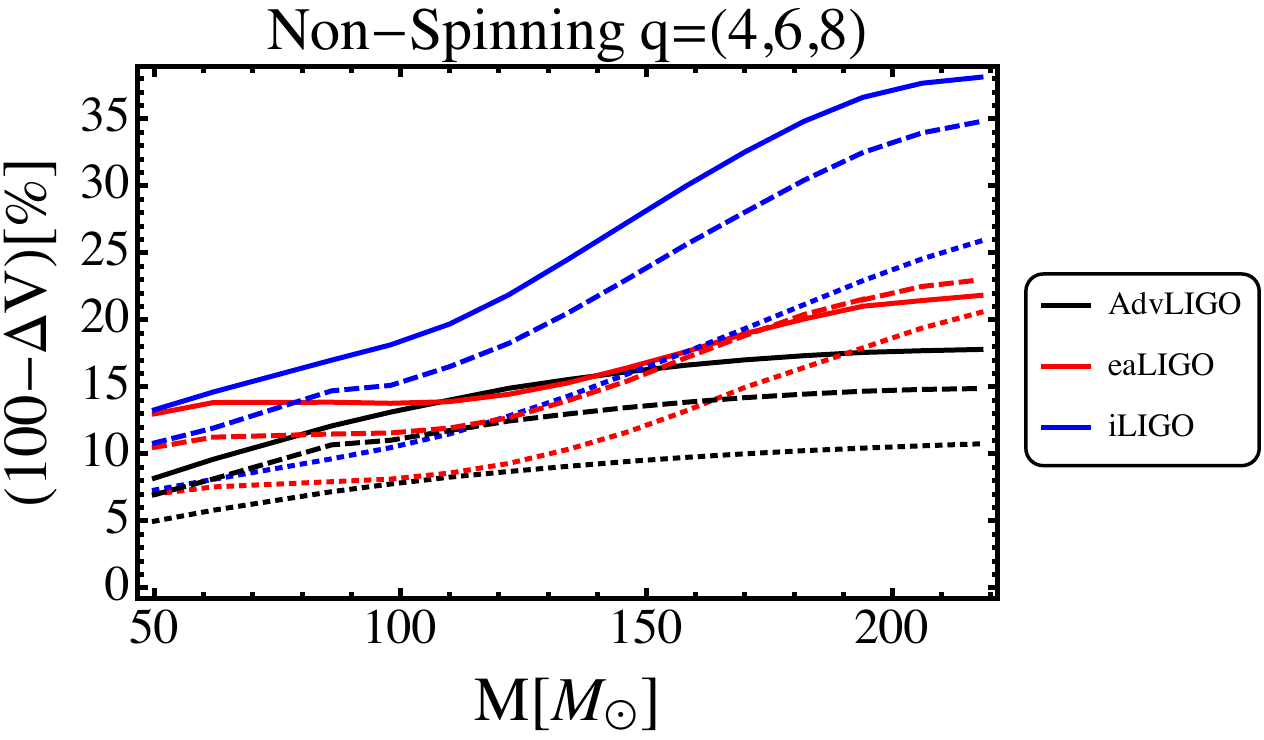}
\includegraphics[width=1\columnwidth]{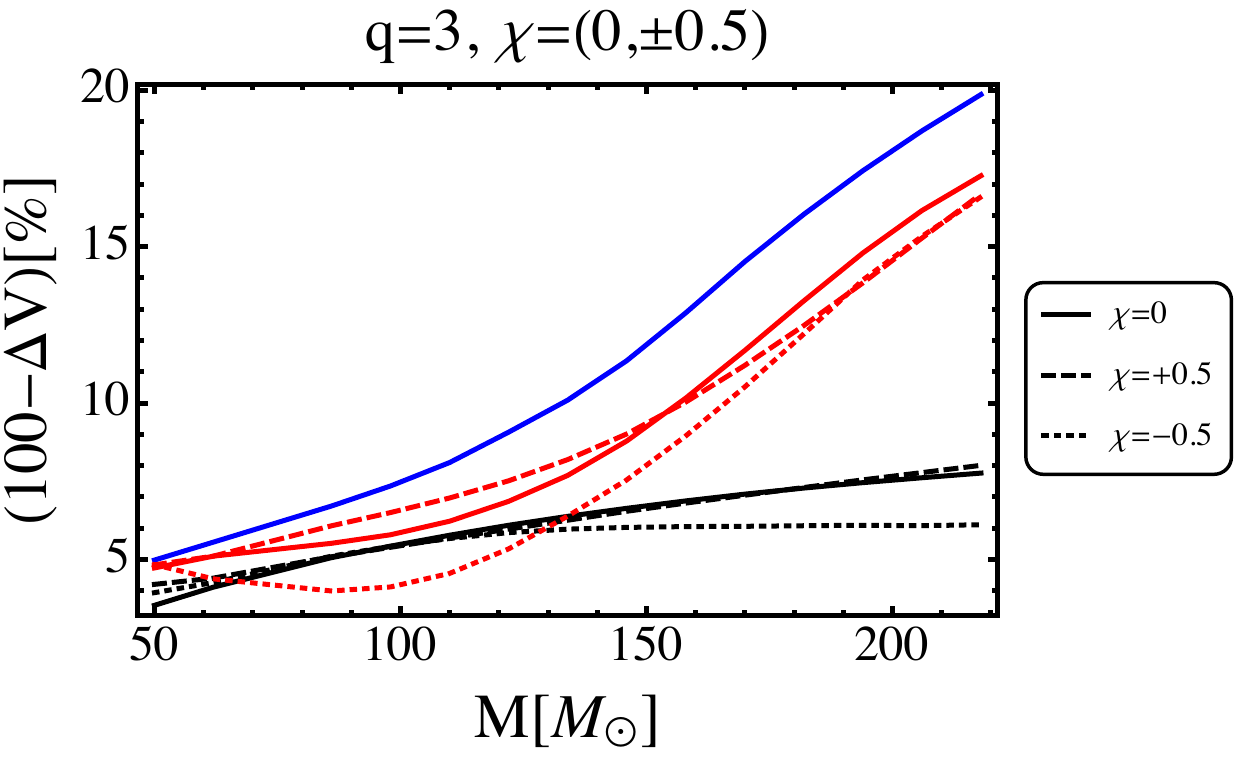}
\caption{Left: Fractional volume loss in $\%$ for non-spinning $q=(4,6,8)$ systems (in dotted, dashed, solid). Rigth: same for $(q;\chi)=(3; 0,\pm 0.5)$. Recall that for the case of iLIGO we have not considered spinning targets.}
\label{ex:fig:volumes}
\end{figure*}
For the bank templates we use an equal-spin $\chi = \chi_1 = \chi_2$ reduced order model (ROM)  \cite{Purrer:2014fza} of SEOBNRv1 \cite{Taracchini:2012ig}. The ROM is constructed in the frequency domain and agrees with SEOBNRv1 waveforms to a mismatch of $< 0.002$ for low mass and $< 0.003$ at high mass. The mismatch can reach $\sim 0.01$ in isolated regions, for very high mass-ratios and/or high anti-aligned spins. This behavior is due to the undersampling of non-quasicircular coefficients in SEOBNRv1. Its range of validity in terms of spin is  $\chi\in[-1,+0.6]$. \\

For each target waveform $h_{i,j}(\Xi_i,\Lambda_j)$ we compute ${\cal F}_{i,j}={\cal F}_{\cal B}h_{i,j}$, the corresponding recovered intrinsic parameters $\Xi^{\cal B}_{i,j}$ and the optimal SNR $\rho_{i,j}=\sqrt{\langle h_{i,j}|h_{i,j} \rangle}$. Maximization of the fitting factor over $\Xi$, is performed running several Nelder Mead Simplex algorithms as implemented in \cite{Mathematica}. We let each of the runs start at different initial regions of the parameter space and the highest result is chosen as the true fitting factor ${\cal F}_{i,j}$. We then compute the fraction of the optimal and suboptimal volumes in which a system $h_i$ with parameters $\Xi_i$ can be detected as 
\begin{equation}
\Delta V [\%]=100 \times R_{i}=100 \times \bigg{(}\frac{\sum_j {\cal F}_{i,j}^3 \rho_{i,j}^3}{\sum_j \rho_{i,j}^3}\bigg{)}
\end{equation}
and the effective fitting factor as ${\cal F}_i^{eff}=R_i^{1/3}$. The observation-averaged recovered parameters are computed as 
\begin{eqnarray}
\Xi_i^{\cal B}=\bigg{(}\frac{\sum_j \Xi_{i,j}^{\cal B} {\cal F}^3_{i,j} \rho^3_{i,j}}{\sum_j  {\cal F}^3_{i,j} \rho^3_{i,j}}\bigg{)}
\end{eqnarray}
and the corresponding averaged parameter bias as
\begin{equation}
\Delta \Xi_i=\Xi_i^{\cal B}-\Xi_{i,0}
\label{eq:PEbiasdef}
\end{equation}
where $ \Xi_{i,0}$ are the recovered parameters for the case that the target waveform does only contain the $(2,2)$ mode. This accounts for intrinsic biases of the template bank towards the quadrupolar modes of our targets and allows to isolate the effect of HM. We note that unlike studies like \cite{Varma:2014jxa}, which quote the absolute value of the parameter bias, we prefer to keep track of its sign, as this can be then compared with a-priory estimates. For instance, since low mass systems have larger frequency content than large mass ones, we expect that the higher mode content of large mass systems will produce averaged-biases to lower masses.\\
In order to asses the significance of these biases, we compare them to the corresponding statistical uncertainty that searches are affected by due to the presence of Gaussian noise in the data. For doing so, we employ the indistinguishability criterion for two waveforms $h$ and $g$ with mismatch $\epsilon=1-{\cal O}[h,g]$ given by \cite{Lindblom:2008cm} and used in \cite{MacDonald:2011ne}. Two waveforms are indistinguishable at a given SNR $\rho$ if $\epsilon<1/2\rho^2 $.  We will thus consider that parameter estimation\footnote{Or measurement following the notation in \cite{Lindblom:2008cm}.} is not compromised due to systematic biases produced by the presence of HM in the target waveform if the best matching template $h^{\cal B}(\Xi_i^{\cal B})$ and the one best matching the injection with no HM $h^{\cal B}(\Xi_{i,0})$ are insdistinguishable. We stress that this method does not provide a complete parameter estimation study, as, for instance, a bayesian MCMC study \cite{vanderSluys:2008qx,Littenberg:2012uj,Graff:2015bba}, would do, but provides a fast first guess of the significance of the systematic parameter bias we find, which we get for free as a result of the fitting factor calculation.
\section{Effect on detection}
In general, as $q$ and $M$ increase, the larger contribution from HM to the target signal makes ${\cal F}^{eff}$ decrease, which is expected from PN theory. For AdvLIGO losses do never reach $20\%$ for any of the studied cases and $10\%$ is reached for high mass $q\geq6$ $M>100M_\odot$ systems. In contrast, mainly due to their higher $f_0$, for both eaLIGO (and iLIGO) losses reach values of $\sim 23\%$ ($\sim 35\%$) for the highest $q$ studied. Losses of $10\%$ occur for all the targets with mass parameters $(q\geq 6, M\geq50M_{\odot})$, except for $q=1$ (which are not shown in any plot due to the negligible losses found) and losses of $20\%$ are present for iLIGO for $q\geq 4$, as can be seen in Fig.\ref{ex:fig:volumes}. The fact that the seismic wall determines the different behavior of eaLIGO and AdvLIGO is clear from the fact that both detectors have similar losses up to masses of $M\sim 110M_{\odot}$, when the $(2,2)$ mode of the target waveform dominates the full signal content in the band of both detectors and can be well filtered by a bank that only contains quadrupolar modes. However, after that point, the $(2,2)$ mode starts to get out of band for eaLIGO  while it remains in for AdvLIGO.\\ 
We note that our predicted losses for AdvLIGO are a a bit lower than those shown in \cite{Varma:2014jxa} due to the inclusion of the effective spin parameter $\chi$ in our template waveforms. This provides an extra degree of freedom that can be exploited by quadrupolar waveforms to filter signals containing HM. This is also the main reason for the different results obtained for iLIGO and eaLIGO.\\
Regarding the effect of spin, no $q=1$ case reached even $3\%$ losses. For the $(q,\chi)=(3,\pm 0.5)$ case, losses are very similar to the ones for $(q,\chi)=(3,0)$ (see Fig.\ref{ex:fig:volumes}, right panel) which is consistent with the statement that spin should be secondary in terms of the impact of HM. Note however, how losses are a bit larger for the positive spin case than for the negative one for low mass. This could be however due to the fact that $\chi=+0.5$ lies in the limit of validity of the SEOBNRv1-ROM model. For high mass, results show that contributions from HM become equally important in terms of ${\cal F}^{eff}$. Furthermore, the losses observed for $\chi=0$ seem a good guess of those observed for the spinning cases, particularly for the highest masses. We note that it would have been interesting to study cases with spins closer to $\pm1$ and higher mass ratios. However, the only case with reasonably high spins and mass ratio available in the SXS catalogue was the $q=3, \chi=\pm0.5$ used here.
\begin{figure}[ht]
\includegraphics[width=.49\columnwidth]{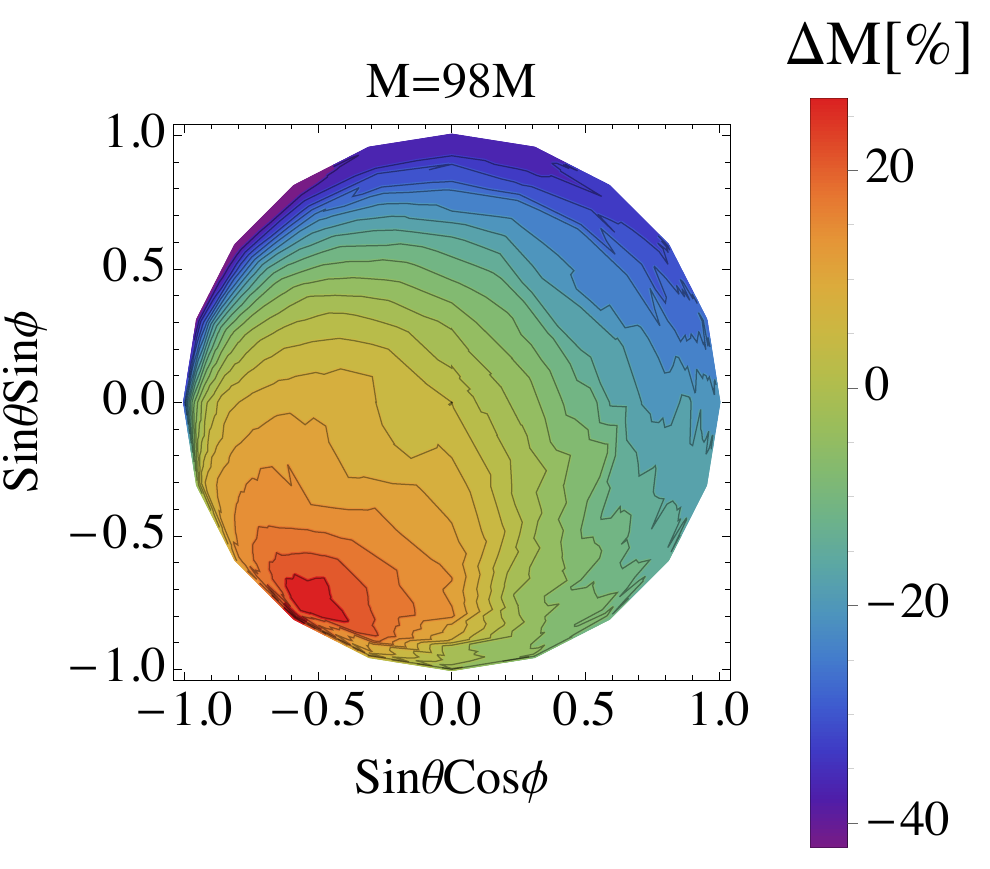}
\includegraphics[width=.49\columnwidth]{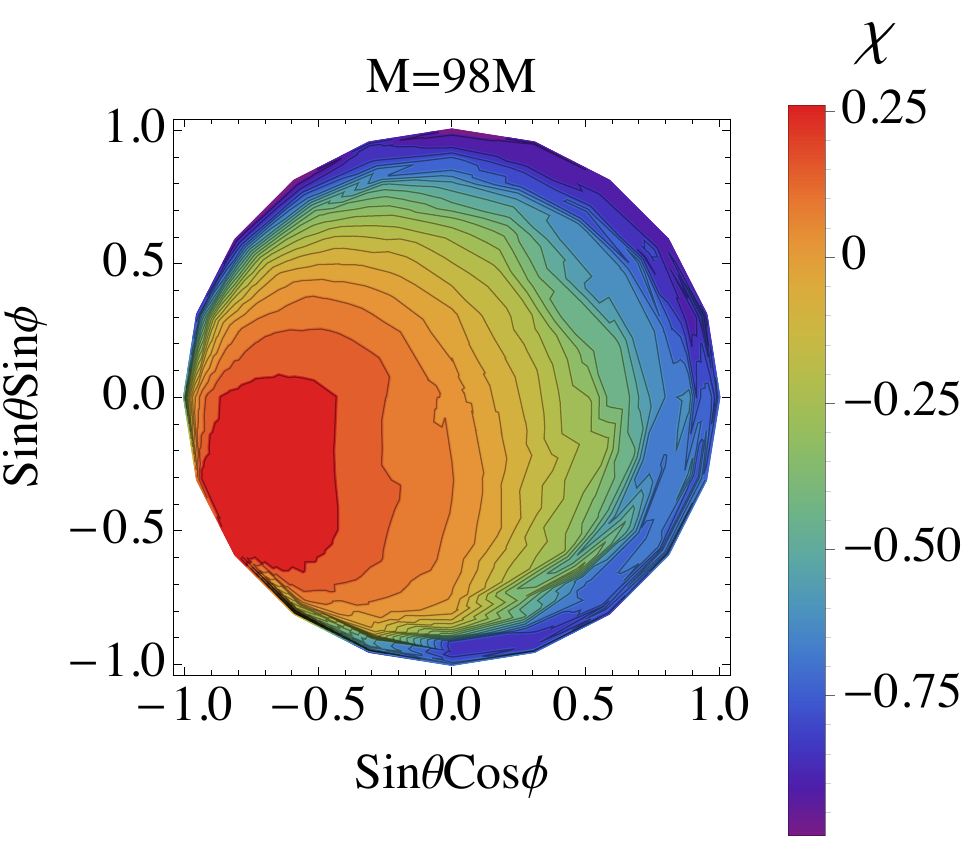}
\includegraphics[width=.49\columnwidth]{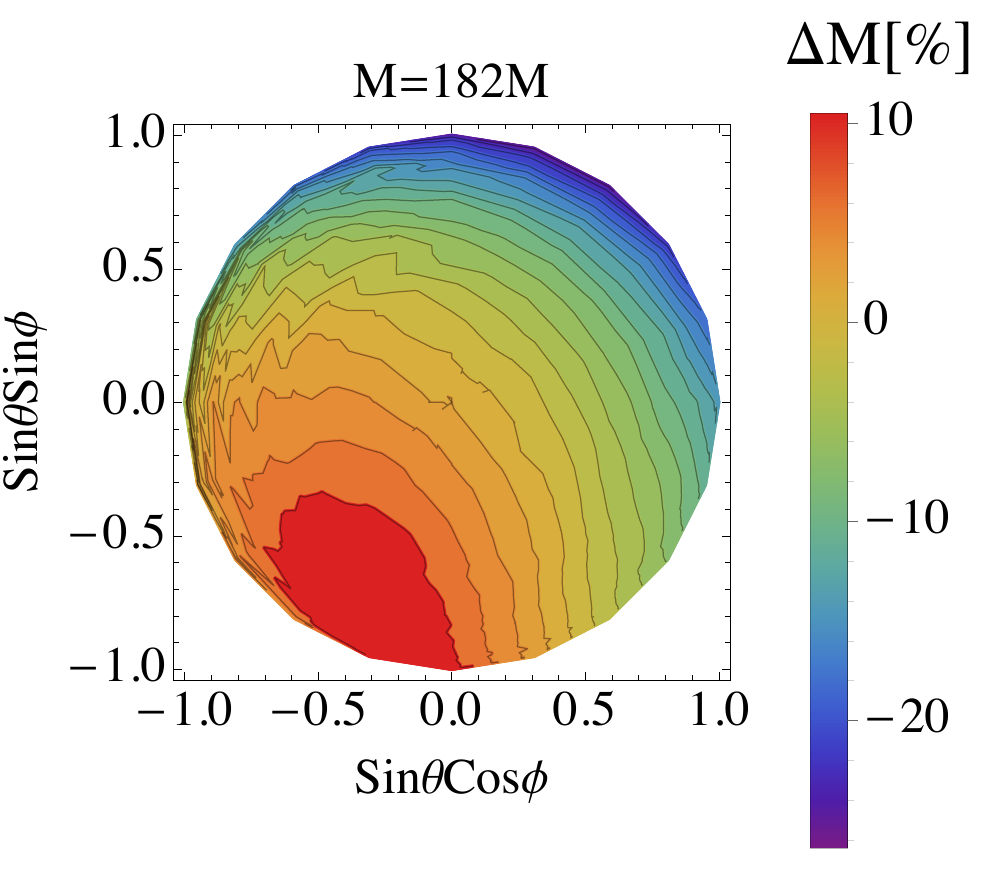}
\includegraphics[width=.49\columnwidth]{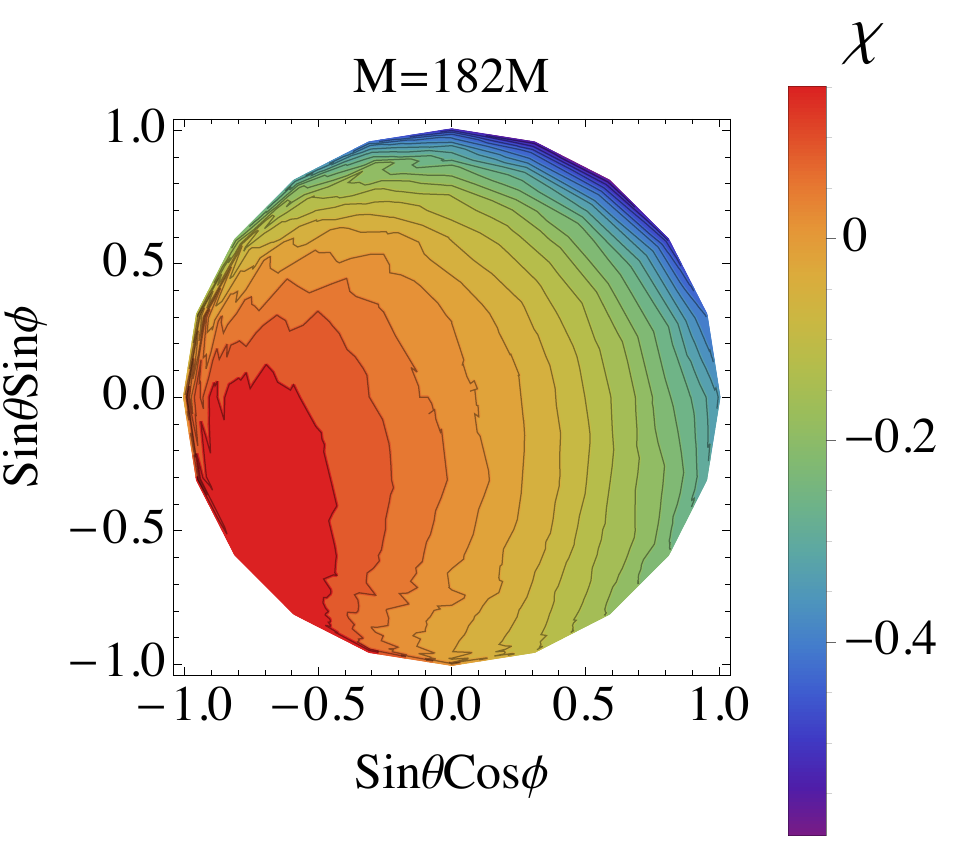}
\caption{Systematic biases obtained for the total mass (left) and effective spin (right) for a $q=3$ non-spinning system for eaLIGO (top) and AdvLIGO (bottom) as a function of the location of the detector on the (upper hemisphere) sky of the source. Note that the different interaction of the modes as a function of the angle $\varphi$ generates biases to either larger or lower values, which in general grow (in absolute value) as $\theta$ does. Biases to low masses are more common due to the higher frequency content of the signal for most values $\varphi$, has to be imitated by low mass templates. Last, note that $\theta=0$ corresponds to the center of the plot while its perimeter corresponds to $\theta=\pi/2$.}
\label{ex:fig:pbx}
\end{figure}
\section{Parameter Bias}

\begin{figure*}[ht]
\includegraphics[scale=.38]{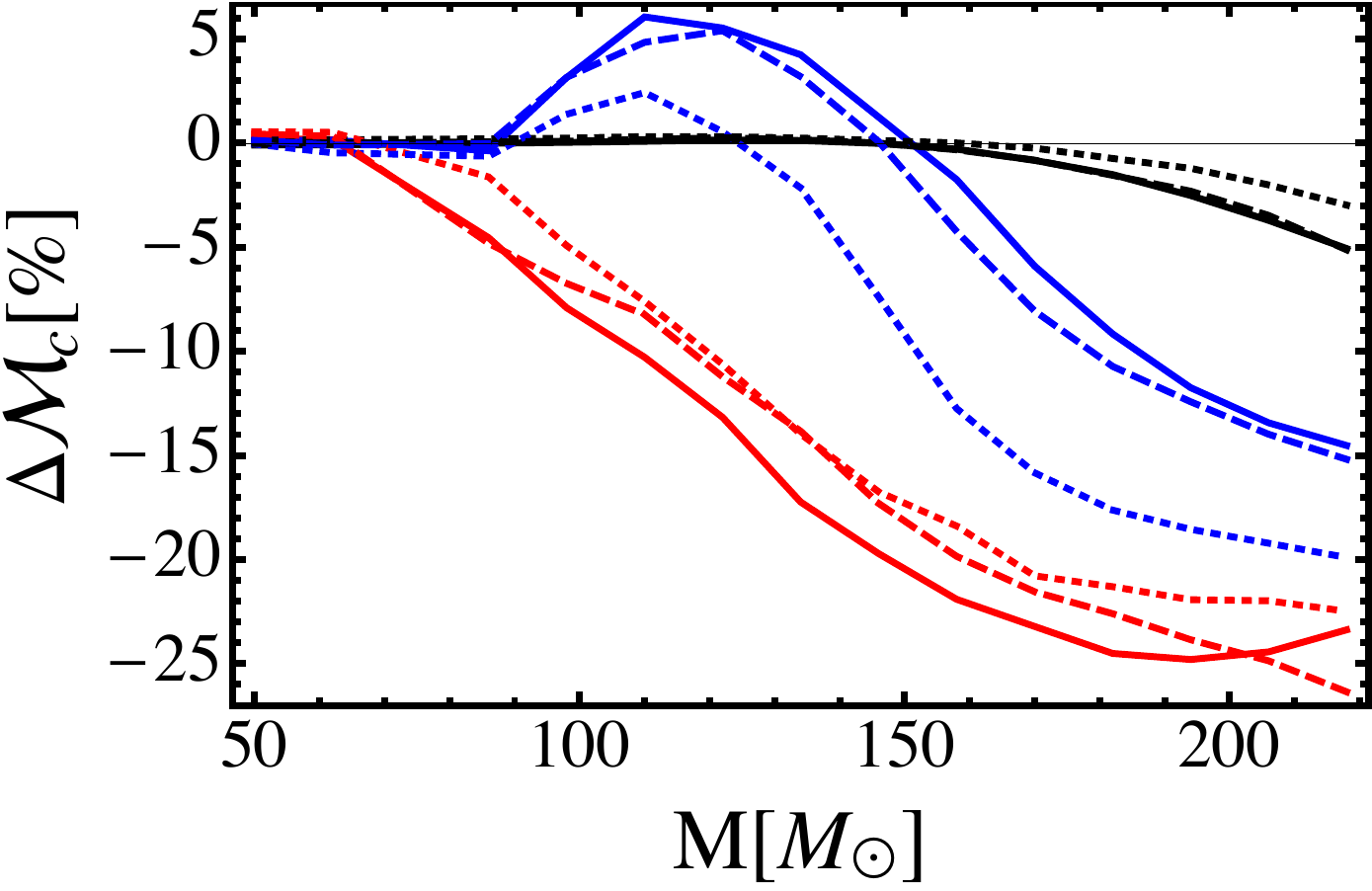}
\includegraphics[scale=.38]{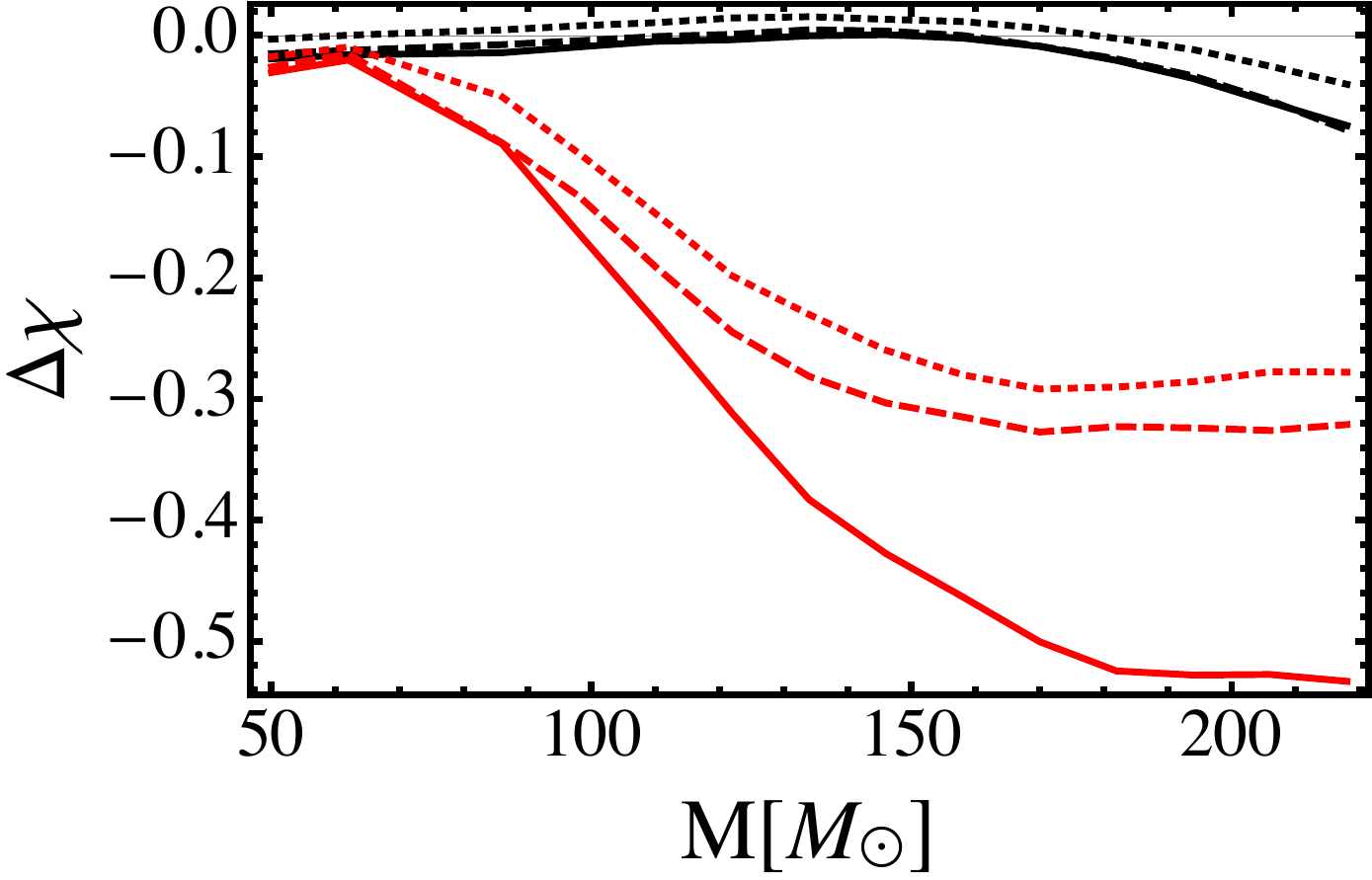}
\includegraphics[scale=.51]{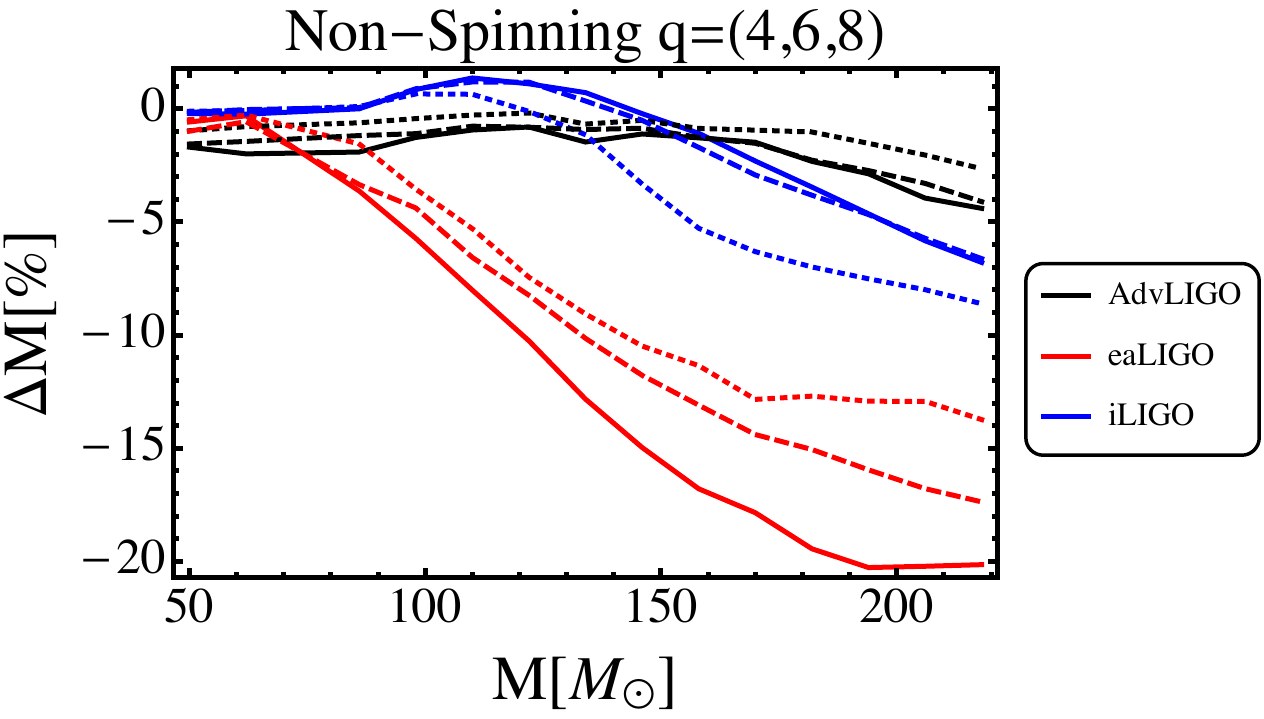}
\includegraphics[scale=.38]{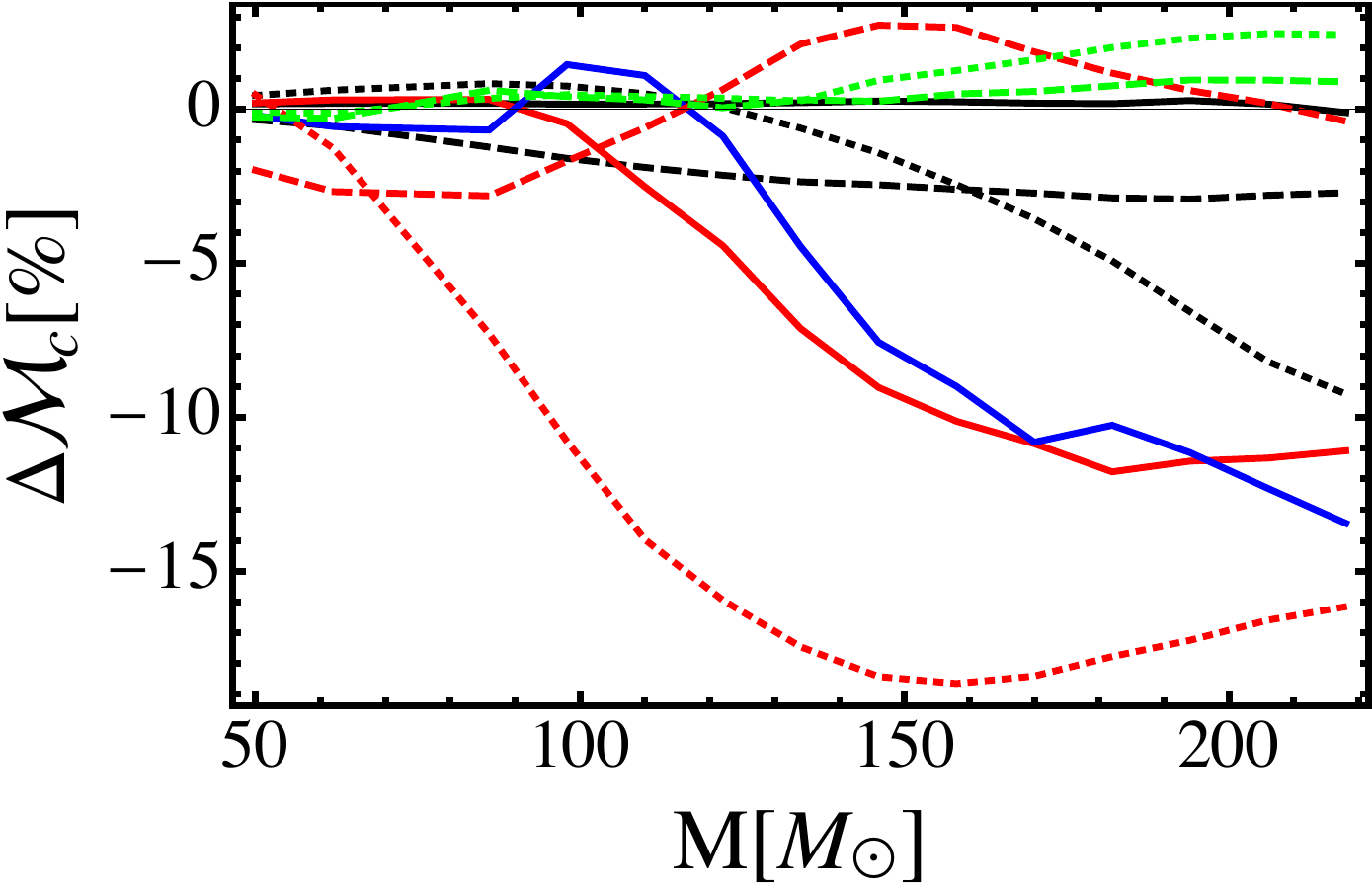}
\includegraphics[scale=.38]{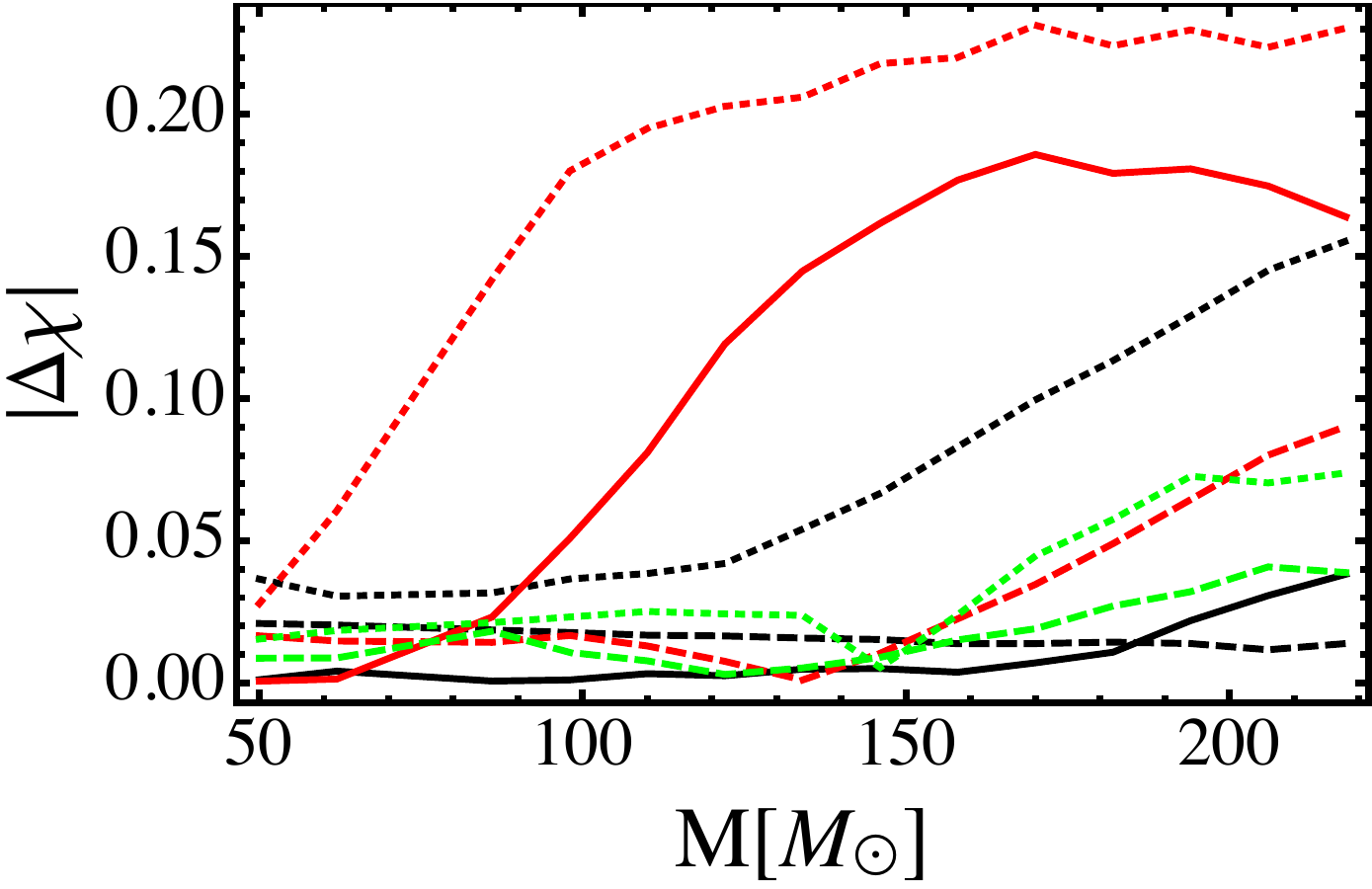}
\includegraphics[scale=.51]{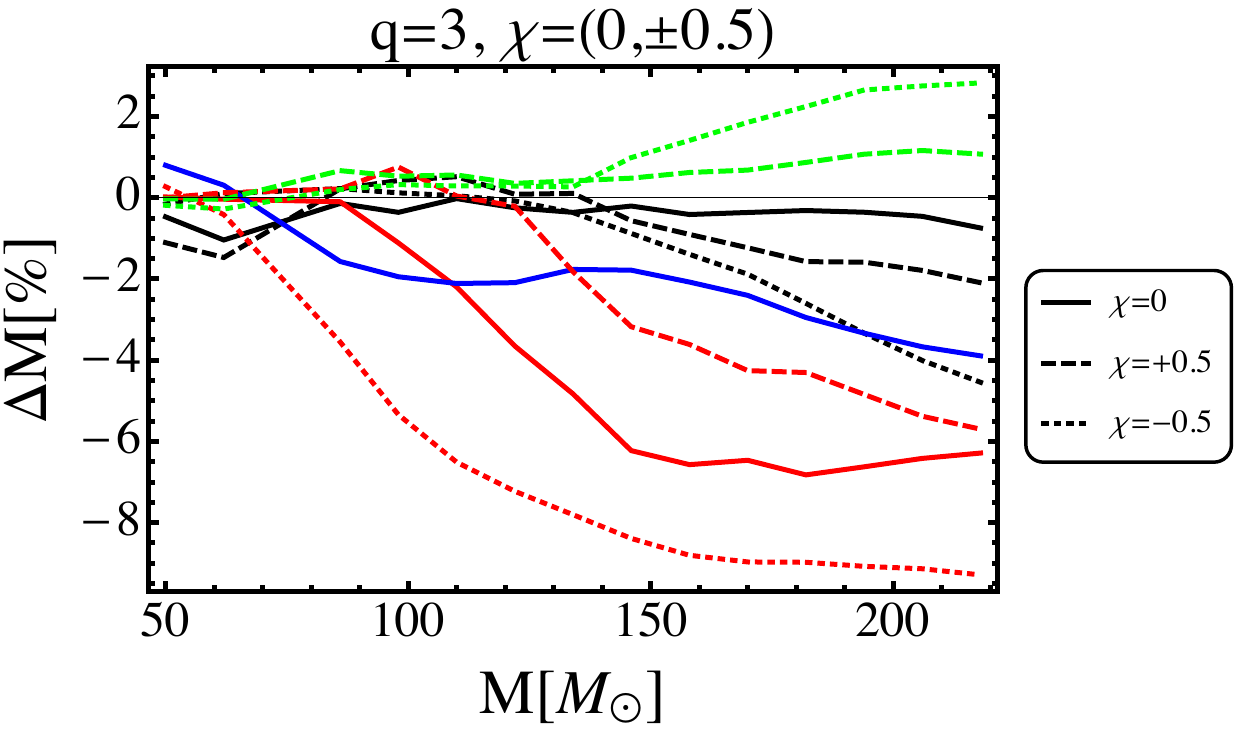}
\caption{Top: $M$, ${\cal M}_c$, and $\chi$ systematic bias for the $q=(4,6,8)$ non-spinning cases. We use the same source and noise curve code as in Fig.\ref{ex:fig:volumes}. Bottom: Same for the $(q;\chi)=(3;0,\pm 0.5)$ in red. We use (dashed,dotted) for (``+'',``-'') spin and add $(q, \chi)=(1,\pm 0.2)$ case in green for eaLIGO.}
\label{ex:fig:pb}
\end{figure*}
Due to its importance in GW data analysis, we will express results not as a function of $(q,M)$ but rather consider the so called chirp mass parameter ${\cal M}_c$ \footnote{${\cal M}_c=M\eta^{3/5}$ $\eta=q/(q+1)^2$} and the total mass $M$. Before discussing the averaged systematic errors measured due to the neglection of HM, we want to note that the intrinsic parameter bias $\Xi_{i,0}$ of the SEOBNRv1-ROM model towards our hybrids containing only the quadrupolar modes were never larger than $(|\Delta M|(\%), |\Delta {\cal M}_c|(\%),|\Delta \chi|) = (2\%,2\%,0.04)$ for all the total mass range, except for the $(q,\chi)=(3,+0.5)$ case, for which these reached maximum values of $(4\%,6\%,0.05)$\footnote{Again, note that $(q,\chi)=(3,+0.5)$ is in the limit of validity of SEOBNRv1-ROM.}.\\
\begin{figure*}[!ht]
\includegraphics[width=1\columnwidth]{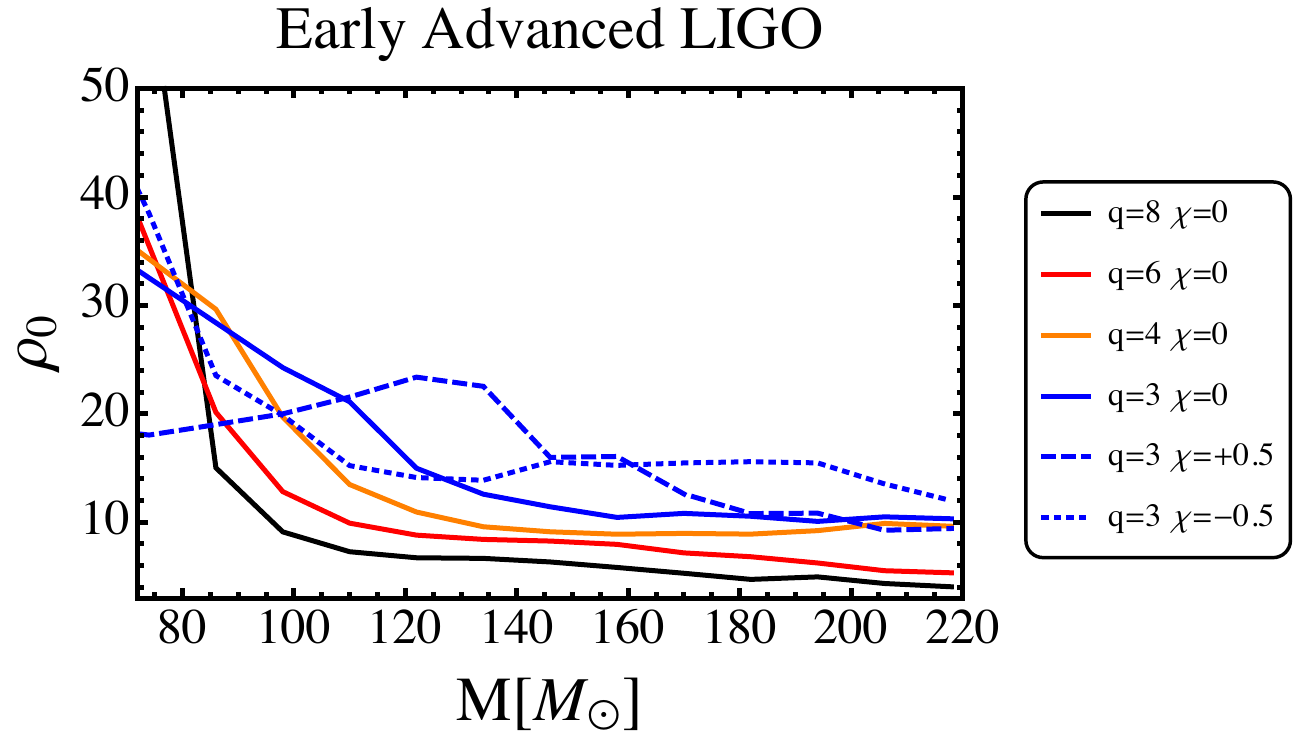}
\includegraphics[width=1\columnwidth]{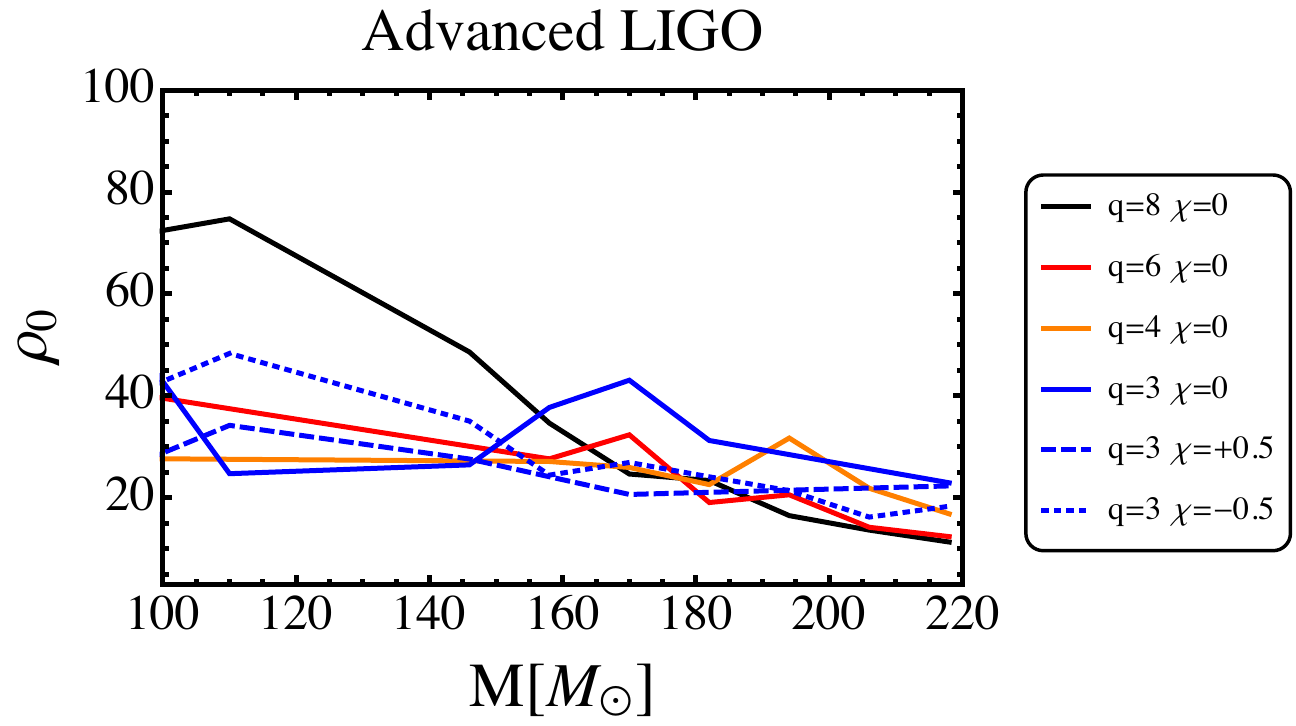}
\caption{Comparison between systematic errors and statistical uncertainties. We show the minimum SNR $\rho_0$ at which systematic biases due to the neglection of HM dominate those due to statistical uncertainties for the studied sources.}
\label{fig:statsys}
\end{figure*}

The main effect of the HM is introducing large frequencies in the detector band, thus one should expect that the quadrupolar SEOBNRv1 waveform best matching a target waveform $h(\Xi)$ with parameters $\Xi$ should have a larger frequency content than that corresponding to the quadrupolar template $h^{\cal B}(\Xi)$ having the intrinsic parameters $\Xi$ of the target. Intuitively, this can be achieved via introducing biases towards lower total mass and larger positive spin. Fig.\ref{ex:fig:pbx} shows the biases in total mass and spin obtained for all values of $(\theta,\varphi)$ (thus averaged over $\psi$) for a $q=3$ non-spinning system for the cases of eaLIGO and AdvLIGO. Note that $\theta=0$ corresponds to the center of the plot while its perimeter corresponds to $\theta=\pi/2$. We see how the two different ways of increasing the template frequency (lowering mass and raising spin) compete along the different $(\theta,\varphi)$. As expected, the absolute value of the bias grows as $\theta$ does. Also, the different interaction of the modes as a function of $\varphi$ generates a sort of dipolar pattern where biases vary from positive to negative. It is remarkable that while averaged biases shown in Fig.\ref{ex:fig:pb} for the systems in Fig. \ref{ex:fig:pbx} are of $(\Delta M, \Delta \chi)\sim (-5\%,-0.1)$ for eaLIGO and $\sim (-3\%, 0)$ for AdvLIGO, biases for particular edge-on orientations can be much larger, up to $(\Delta M, \Delta \chi) \sim (-40\%, -0.7)$ for the case shown for eaLIGO and  $\sim (-20\%, -0.4)$ for the one shown for AdvLIGO. Note also that even though the total mass chosen for the eaLIGO example is almost a half of that chosen for AdvLIGO, systematic biases are much lower for the latter case due to the lower $f_0$ of AdvLIGO, which makes it much more sensitive to the long PN inspiral dominated by the quadrupolar modes.\\
Fig.~\ref{ex:fig:pb} shows the averaged parameter bias over the observable volume, given by Eq.\eqref{eq:PEbiasdef}, for the studied targets. As a general trend, neglection of HM causes observation-averaged biases towards lower ($\chi$, $M$, ${\cal M}_c$) which increase as $M$ and $q$ do. As expected, biases are much larger for iLIGO and eaLIGO than for Adv.LIGO. In particular, note that the lower $f_0$ of Adv.LIGO allows for an excellent recovery  of ${\cal M}_c$ for most of the $M$ range. This is due to the larger weight of the PN inspiral in the detector band. Regarding spinning cases, systematic biases are larger for negative spin cases than for positive spin ones. For $q=1$ we only show the eaLIGO cases, which were the only ones having systematic biases comparable to those of the other cases.\\

We now compare the observation-averaged biases to the statistical uncertainty we expect for each detector via computing the minimum SNR $\rho_{0}$ at which PE would be dominated by the systematic biases. We note that, unlike the volume loss $R_i$, the quantity $\rho_0=\sqrt{1/2\epsilon}$ is extremely sensitive to tiny variations in the parameters recovered by the Nelder-Mead algorithm, which has the risk of settling in a local maximum. In particular, for an error $\Delta \epsilon$ in the estimation of $\epsilon$, one gets a variation for $\rho_0$ of $\Delta \rho_0\sim \epsilon^{-3/2} \Delta \epsilon$. This will specially affect regions of the parameter space where systematic biases are lower and where the parameter space is denser \footnote{Where presumably $\epsilon$ will be lower and $\Delta \epsilon$ will be larger.}: so for low mass, large mass ratio, positive spin and AdvLIGO. Due to this, although we run up to 15 times some of the Nelder-Meads, Fig. \ref{fig:statsys} shows several peaks that do only allow us to give a rough estimate of $\rho_0$. Also, for the same reason, for AdvLIGO we only show results for $M\geq 100M_\odot$. Results suggest that for AdvLIGO, PE at SNR $\rho \simeq 8$ would be affected by HM for $M\geq 220M\odot$.  However, for the case of eaLIGO, this limit gets reduced to $M\geq 100M_\odot$ due to the larger systematic biases. 

\subsection{On the usage of the SEOBNRv2 waveform model.}
We note that during this study, the SEOBNRv2 ROM waveform model \cite{Taracchini:2012ig,Taracchini:2013rva,Purrer:2014fza} became available. This model does not only supersede SEOBNRv1 ROM but also covers a wider spin range, namely $\chi_{v2}\in [-1,+1]$ while $\chi_{v1}\in [-1,+0.6]$. For this reason, we suspected that qualitatively, our results for the $(q=3,\chi=+0.5)$ case might be different when using SEOBNRv2 as quadrupolar model.
As a sanity check, we re-computed the event loss and parameter bias using the SEOBNRv2-ROM family as quadrupolar template model for the cases of the non-spinning $q=8$ target and for the $q=3$, $\chi=+0.5$ one. Note that the latter is close to the limit of validity of SEOBNRv1-ROM but well inside the one of SEOBNRv2-ROM. Qualitatively, both models yielded the same trend in terms of event loss and parameter bias: larger losses as $q$ and $M$ increase and observation-averaged biases towards lower $q$, $M$ and $\chi$. Quantitatively, both studies (using v1 and v2) yielded very similar results. The exception to this was the bias of the chirp mass, which differed by up to a $50\%$ for both the $q=8$ and the $q=3$ cases when considering the eaLIGO noise curve. Also,  SEOBNRv1 shown to be better at recovering the HM content of the spinning system. Since none of the models are expected to model HM, we don't find any reason why we should expect the converse to happen.
\section{Conclusions}
In this paper we have studied the impact of the current neglection of HM in GW searches for binary black holes. We have extended previous studies, which focused in non-spinning searches, non-spinning target signals and AdvLIGO to the case of single-aligned spin searches and targets and to the case of AdvLIGO and the upcoming eaLIGO. We have also considered the case of a non-spinning search and targets for the case of iLIGO. The main results of this study are the following.
\begin{itemize}
\item Including an effective spin parameter $\chi$ in our template bank reduces the losses observed in \cite{Varma:2014jxa} and \cite{Capano:2013raa} for the case of non-spinning targets.
\item The higher frequency cutoff $f_0$ of the upcoming eaLIGO, makes losses due to the neglection of HM significantly larger than those obtained for AdvLIGO in \cite{Varma:2014jxa} and \cite{Capano:2013raa} (even though we use an aligned-spin bank) and increases the region of the parameter space where HM are needed. The same applies for iLIGO.
\item When an aligned spin template bank is used, neglection of HM leads in general to observation-averaged biases to lower total mass, chirp mass and spin. These are much larger for the case of eaLIGO and iLIGO due to their larger $f_0$. In particular, for non-spinning targets, biases up to $\Delta \chi=-0.5$ can be obtained for eaLIGO.
\item Losses for spinning systems are quite similar to those for non-spinning systems, the mass ratio and total mass being the dominant parameters.
\end{itemize}
In more detail, we have shown that when an effective spin parameter is included in the template bank, neglection of HM in CBC searches is likely to generate losses $>10\%$ for the $q\geq6$,$M\geq100M_\odot$ regions of the explored parameter space in the case of AdvLIGO. This region is tinier than that obtained in \cite{Varma:2014jxa} ($q\geq4$), due to the fact that they used a non-spinning template bank. However, for the case of eaLIGO (and a non-spinning search for iLIGO) we have found potential losses of up to $23\%$ $(39)\%$ due to such a neglection. Losses of $10\%$ happen for eaLIGO for the $q\geq4$, $M\geq150M_\odot$ and $M>50_\odot$ for $q\geq6$. Furthermore, for the eaLIGO case, averaged systematic biases affecting parameter estimation are normally above $(\Delta M, \Delta \chi, \Delta \mathcal{M})=(-5\%,-10\%,-0.1)$ for the most part of the explored parameter space and reach values of $(-15\%,-25\%,-0.5)$ for the highest $(q,M)$ cases. We compared the systematic biases to the corresponding statistical uncertainties. Results for eaLIGO suggest that measurements with SNR$\simeq 8$ would be affected by the presence of HM at $M\geq 100M_\odot $ for the largest $q$ considered. In the case of AdvLIGO, we estimate that PE is likely to be affected at $\rho\sim 8$ for $M \geq 220_\odot$ for the largest $q$ studied. These value is larger than that obtained by  \cite{Varma:2014jxa}, however comparing the two results is intricate since they used non-spinning templates. \\
The study of the FAR of a GW search including higher modes is out of the scope of this work. This is however is a crucial instrument for assessing the real significance of the losses we find and for assessing the need of such a search. Capano et al., \cite{Capano:2013raa} demonstrated that the threshold SNR needed for claiming a trigger would have to be raised by roughly $10\%$ due to the larger number of templates needed for such a search, which roughly means that the event losses of  a search non-including HM w.r.t., a one including them would roughly be $90\%$ of those obtained in this paper. Also, this paper has not considered the effect of signal-based vetoes as the $\chi^2$ \cite{Allen:2004gu}, used in GW searches \cite{Babak:2012zx} for discriminating real signals from background noise transients, known as glitches. This would especially punish signals for which we found poor fitting factors (which would be treated as glitches), leading to larger event losses. An obvious limitation of this work is the low number of spinning cases considered. This is due to the lack of public aligned spin NR waveforms with high HM content. We chose for this study an SXS case where HM were expected to be weak $(q=1,\chi=\pm 0.2)$ and the one were HM where the strongest possible while having equal spins $(q=3,\chi=\pm 0.5)$. We aim to extend this study to general unequal spin targets and unequal spin template bank. We end pointing that another interesting extension of this work would be to consider the case of precessing targets. 
\section{Acknowledgements}
We thank Alejandro Boh\'{e} for many useful discussions, Francisco Jim\'{e}nez for fruitful discussions regarding parameter biases and Alex Nielsen for comments on the manuscript.
We are grateful to the SXS Collaboration for the development and public release of  their NR catalog, on which our hybrid target waveforms rely. JCB, SH and AMS are supported by the Spanish Ministerio de Econom\'ia y Competitividad (grants FPA2013-41042-P  and CSD2009-00064) and the Conselleria d'Innovaci\'o, Recerca i Turisme of the Govern de les Illes Balears. JCB also acknowledges support from the Max Planck-Prince of Asturias Mobility Award and thanks the hospitality of AEI Hannover, where part of this paper was written. MP was supported by Science and Technology Facilities Council grant ST/I001085/1. This paper has LIGO DCC number LIGO-P1500184.

\section*{Appendix 1: the unequal spin case $(q=8,\chi_1=0,\chi_2=-0.5)$}
\begin{figure}[!ht]
\centering
\includegraphics[width=1\columnwidth]{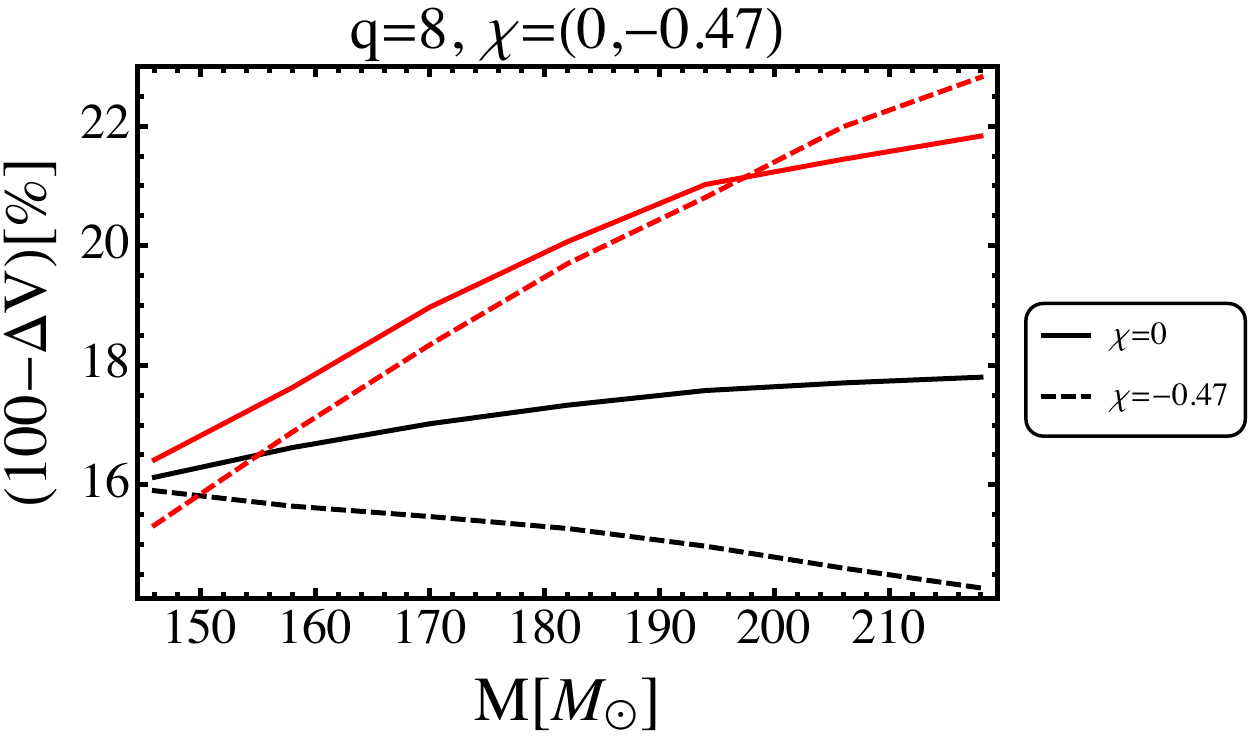}
\caption{Fractional volume loss in $\%$ a for $q=8$  non-spinning system and an unequal spin source with dimensionless spins $(\chi_1,\chi_2)=(0 ,-0.5)$.}
\label{ex:fig:volumesu}
\end{figure}
In order to further test the validity of our statement that the effect of spin on the HM contribution should be secondary in terms of detection losses, we studied the case of the unequal spin case $(q=8,\chi_1=0,\chi_2=-0.5)$, where the spinning black hole is the ``heavy one''. Note that the quadrupolar modes of this source are not expected to be well modelled by the SEOBNRv1-ROM single spin model. However, the effective spin $\chi$ can be computed for this system, obtaining $\chi=-4/9\simeq -0.47$. Fig. \ref{ex:fig:volumesu} shows the sky-averaged event losses obtained from the effectualness of the SEOBNRv1 ROM model to the unequal spin system and those corresponding to the $q=8$ non spinning system. 

We note that for this case, we have not used a hybrid PN/NR waveform as a target. Instead we have used a NR waveform. For this reason the plot starts at $M=150M_\odot$, which corresponds to the lower mass at which the NR simulation reaches the $10$Hz lower frequency cutoff of  AdvLIGO. Once again it can be noted that losses are very similar for the two cases: for eaLIGO they are basically the same and for AdvLIGO losses are a bit lower for the (negative) spin case, as it happens for the $q=3$ cases shown in Fig.\ref{ex:fig:volumesu}. This reinforces our statement that spin has a secondary contribution to the strength of HM.

\bibliography{HMbib}

\end{document}